\documentclass[sigconf]{acmart}

\usepackage{type1cm}    
\usepackage{enumerate}
\usepackage{enumitem}
\usepackage{graphicx}
\usepackage{epstopdf}
\usepackage{epsfig}
\usepackage{color}
\usepackage{multirow}
\usepackage{etoolbox}
\usepackage{booktabs}
\usepackage{amsmath}
\usepackage{amssymb}
\usepackage{mathtools}
\usepackage{hhline}
\usepackage{setspace}
\usepackage{balance}
\usepackage{xcolor,colortbl}
\usepackage{array}
\usepackage{xspace}     
\usepackage{balance}     
\usepackage[font={small}]{caption}     
\usepackage{caption}
\usepackage{siunitx}    
\usepackage[vlined,linesnumbered,ruled,noend]{algorithm2e}
\usepackage{lscape}
\usepackage{subfigure}
\usepackage{array}
\usepackage{pifont}
\usepackage[labelfont=bf]{caption}
\usepackage{alltt}
\usepackage{listings}
\usepackage{multirow}
\usepackage{hyperref}
\usepackage{makecell}

\usepackage{url}

\usepackage{breakurl}

\copyrightyear{2019}
\acmYear{2019} 
\setcopyright{iw3c2w3}
\acmConference[WWW '19]{Proceedings of the 2019 World Wide Web Conference}{May 13--17, 2019}{San Francisco, CA, USA}
\acmBooktitle{Proceedings of the 2019 World Wide Web Conference (WWW '19), May 13--17, 2019, San Francisco, CA, USA}
\acmPrice{}
\acmDOI{10.1145/3308558.3313542}
\acmISBN{978-1-4503-6674-8/19/05}

\newcommand{\csIDs}{22329}
\newcommand{\csNumber}{263635}
\newcommand{\cs}{Cookie Synchronization}

\newcommand{\csync}{CSync}
\newcommand{\toolname}{CONRAD}
\newcommand{\csyncs}{CSyncs}

\newcommand{\totalUsers}{850}

\newenvironment {squishlist}
{\begin{list}{$\bullet$}
  { \setlength{\itemsep}{1pt}
     \setlength{\parsep}{1pt}
     \setlength{\topsep}{1pt}
     \setlength{\partopsep}{1pt}
     \setlength{\leftmargin}{1.5em}
     \setlength{\labelwidth}{1em}
     \setlength{\labelsep}{0.5em} } }
{\end{list}}

\def\ttfntsize{9.5}
\makeatletter
\let\oldtexttt\texttt
\let\texttt\@undefined
\makeatother
\newcommand{\texttt}[1]{\fontsize{\ttfntsize}{\ttfntsize}\oldtexttt{#1}}
\makeatletter
\let\oldtt\tt
\let\tt\@undefined
\makeatother
\newcommand{\tt}{\fontsize{\ttfntsize}{\ttfntsize}\oldtt}

\definecolor{heraldBlue}{rgb}{0.0,0.0,0.8}



\begin{document}

\title{\cs: Everything You Always Wanted to Know But Were Afraid to Ask}

\author{Panagiotis Papadopoulos}
\affiliation{
	\institution{Brave Software}
}
\email{panpap@brave.com}

\author{Nicolas Kourtellis}
\affiliation{
	\institution{Telefonica Research, Spain}
}
\email{nicolas.kourtellis@telefonica.com}

\author{Evangelos P. Markatos}
\affiliation{
	\institution{FORTH-ICS, Greece}
}
\email{markatos@ics.forth.gr}


\begin{abstract}
User data is the primary input of digital advertising, fueling the free Internet as we know it.
As a result, web companies invest a lot in elaborate tracking mechanisms to acquire  user data that can sell to data markets and advertisers.
However, with same-origin policy and cookies as a primary identification mechanism on the web, each tracker knows the same user with a different ID.
To mitigate this, \cs\ (\csync) came to the rescue, facilitating an information sharing channel between 3rd-parties that may or not have direct access to the website the user visits.
In the background, with \csync, they merge user data they own, but also reconstruct a user's browsing history, bypassing the same origin policy.

In this paper, we perform a first to our knowledge in-depth study of \csync\ in the wild, using a year-long weblog from \totalUsers\ real mobile users.
Through our study, we aim to understand the characteristics of the \csync\ protocol and the impact it has on web users' privacy.
For this, we design and implement \toolname, a holistic mechanism to detect \csync\ events at real time, and the privacy loss on the user side, even when the synced IDs are obfuscated.
Using \toolname, we find that $97\%$ of the regular web users are exposed to \csync: most of them within the first week of their browsing, and the median userID gets leaked, on average, to 3.5 different domains.
Finally, we see that \csync\ increases the number of domains that track the user by a factor of 6.75.
\end{abstract}

\maketitle


\section{Introduction}
\label{sec:introduction}
In the online era, where behavioural advertising fuels the majority of Internet, user privacy has become a commodity that is being bought and sold in a complex, and often ad-hoc, data ecosystem~\cite{radioshakData,brokserSell,toysmartData,Gill:2013:BPF:2504730.2504768}.
Users' personal data collected by IT companies constitute a valuable asset, whose quality and quantity significantly affect each company's overall market value~\cite{businessValuation}.
As a consequence, it is of no doubt that in order to gain advantage over their competitors, web companies such as advertisers and trackers participate in a user data collecting spree, aiming to retrieve as much information as possible and form user profiles.
These detailed profiles contain personal data~\cite{pii2018} such as interests, preferences, personal identifying information, geolocations, etc.~\cite{Papadopoulos:2017:LPD:3038912.3052691,carrascosa2015always,vallina2016tracking}, and could be sold to 3rd-parties for advertising or other purposes beyond the control of the user~\cite{radioshack,facebookCS1,facebookCS2}.

Highlighting the importance of this data collection, web companies have invested a lot in elaborate user tracking mechanisms.
The most traditional one includes the use of cookies: they have been commonly used in the Web to save and maintain some kind of state on the web client's side.
This state has been used as an identifier to authenticate users across different sessions and domains.
Initially, \emph{1st-party} cookies were used to track users when they repeatedly visited the same site, and later, \emph{3rd-party} cookies were invented to track users when they move from one website to another.
The same-origin policy (SOP) was invented a few years later~\cite{sameOrigin}  to restrict the potential amount of information trackers can collect about a user and share with other 3rd-party platforms.

To overcome this restriction, and create unified identifiers for each user, the ad-industry invented the \cs\ (\csync) process~\cite{googleCS}: a mechanism that can practically ``circumvent'' the same-origin policy, and allow web companies to share (synchronize) cookies, and match the different IDs they assign for the same user while they browse the web.
Sadly, recent results show that most of the 3rd-parties are involved in \csync: 157 of top 200 websites (i.e. 78\%) have 3rd-parties which synchronize cookies with at least one other party, and they can reconstruct 62-73\% of a user's browsing history~\cite{Englehardt:2016:OTM:2976749.2978313}.
Furthermore, 95\% of pages visited contain 3rd-party requests to potential trackers and 78\% attempt to transfer unsafe data~\cite{Yu:2016:TT:2872427.2883028}.
Finally, a mechanism for respawing cookies has been identified, with consequences in the reconstruction of users' browsing history, even if they delete their cookies~\cite{Acar:2014:WNF:2660267.2660347}. 

Although past works highlight the use of \csync\ across the most popular sites worldwide, they avoid diving into the details and fail to thoroughly explore this increasingly popular technique.
In fact, the majority of related works perform their analysis using crawled data by fetching the top Alexa websites.
Consequently, little is known on how \csync\ works in the wild, how many of real users' cookies are getting synced during their everyday browsing, and to what extend it affects the end users' privacy.
Importantly, existing studies focus solely on desktop web; however, today more than 52.2\% of all web traffic is generated through mobile phones (up from 50.3\% in the previous year~\cite{mobilePerc}).
This traffic points to a whole new, unexplored ecosystem of mobile Web, where users browse through their very personal devices, while employing a variety of sensors to experience a highly personalized content, but always at the expense of privacy.
Thus, what is the impact of \csync\ in this new mobile ecosystem?
\emph{Which are the basic characteristics of \csync, and the mechanics used?}
\emph{How does \csync\ impact the user's privacy and anonymity on the mobile web?}

In this paper, we aim to answer these questions, by studying \cs\ using a large, year-long dataset of \totalUsers\ real mobile users, and exploring in depth its use and growth through time, dominant companies, along with its side-effects on user privacy.
The contributions of this work are as follows:

\begin{squishlist}
\item We design and implement \toolname: (COokie syNchRonizAtion Detector) a holistic mechanism to detect at real time \csync\ events and the privacy loss on the user side, even when synced IDs are concealed or obfuscated by participating companies aiming to reduce identifiability from traditional, heuristic-based detection algorithms.
In such cases, when non-ID related features are used, our approach achieves high accuracy (84\%-90\%) and AUC (0.89-0.97).

\item We perform the first of its kind, large-scale, longitudinal study of \cs\ on mobile users in the wild.
We perform a passive data collection of the activity of \totalUsers\ volunteering mobile users, lasting an entire year.
This means that the data collected are not crawled, like in past studies, and therefore do not capture a distorted or biased picture of \csync\ on the Web.
Instead, these data provide a rare glimpse of \csync\ in the mobile space, and an opportunity to study \csync\ and its impact on real mobile users' privacy and anonymity.

\item Using the proposed detection mechanism, we conduct an in-depth privacy analysis of \csync.
Our results show that 97\% of regular web users are exposed to \csync.
In addition, the average user receives $\sim$1 synchronization per 68 HTTP requests, and the median userID gets leaked, on average, to 3.5 different domains.
Furthermore, \csync\ increases the number of domains that track the user by a factor of 6.75.
Finally, we detect \csync\ involved in different scenarios such as breaking-off an SSL session, and exposing userIDs and other personal data in cleartext.
\end{squishlist}


\section{Cookie Synchronization}
\label{sec:cookieSync}

\subsection{How does \cs\ work?}
\begin{figure}[t]
	\centering
	\includegraphics[width=0.73\linewidth]{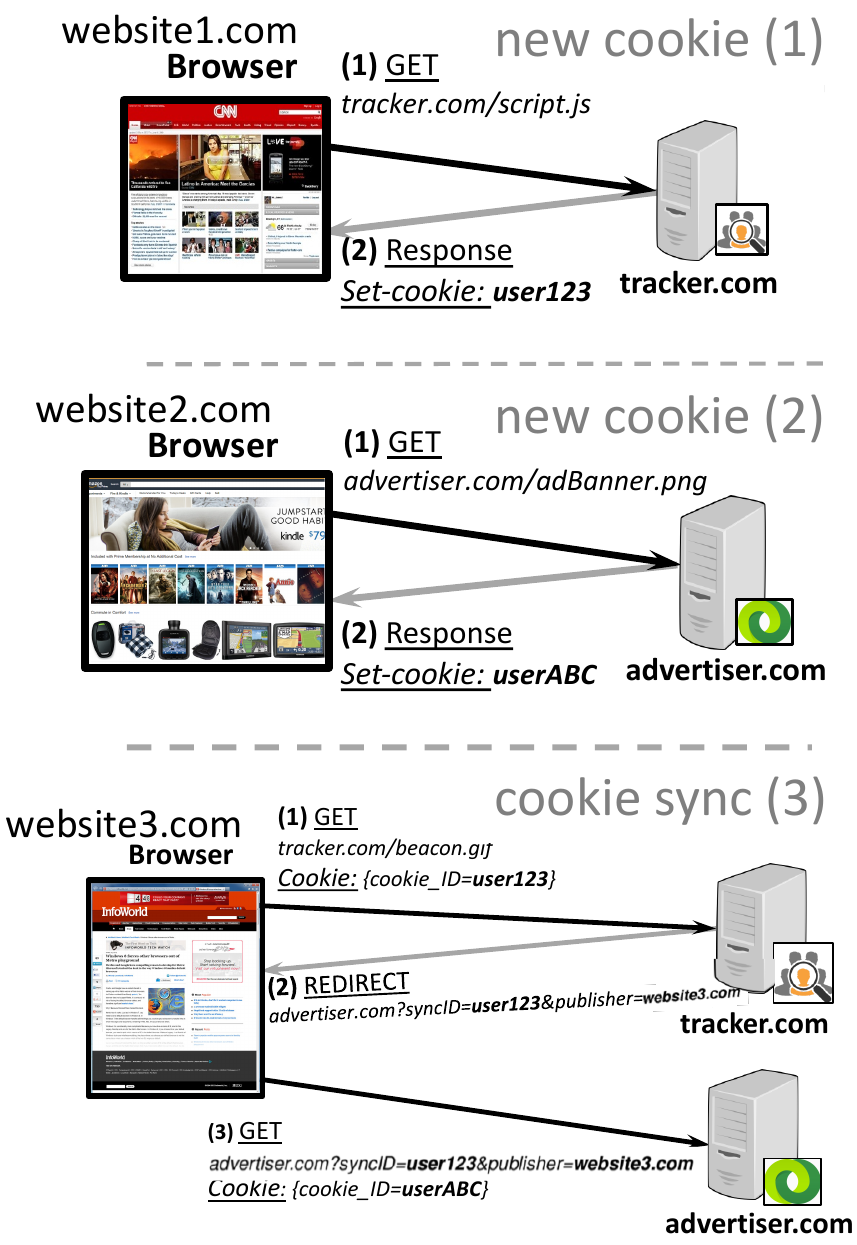}\vspace{-0.5cm}
	\caption{Example of advertiser.com and tracker.com synchronizing their cookieIDs. Interestingly, and without having any code in website3, advertiser.com learns that: (i) cookieIDs userABC==user123 and (ii) userABC has just visited the given website. Finally, both domains can conduct server-to-server user data merges.}
	\label{fig:csync}
\end{figure}

Figure~\ref{fig:csync} presents a simple example to understand in practice what is \csync\ and how it works.
Let us assume a user browsing several domains like {\tt website1.com} and {\tt website2.com}, in which there are 3rd-parties like {\tt tracker.com} and {\tt advertiser.com}, respectively. 
Consequently, these two 3rd-parties have the chance to set their own cookies on the user's browser, in order to re-identify the user in the future.
Hence, tracker.com knows the user with the ID {\tt user123}, and advertiser.com knows the same user with the ID {\tt userABC}.
Now let us assume that the user lands on a website (say {\tt website3.com}), which includes some JavaScript code from tracker.com but not from advertiser.com.
Thus, advertiser.com {\bf does not (and cannot) know which users visit website3.com}.
However, as soon as the code of tracker.com is called, a GET request is issued by the browser to tracker.com (step 1), and it responds back with a REDIRECT request (step 2), instructing the user's browser to issue another GET request to its collaborator advertiser.com this time, using a specifically crafted URL (step 3):

\vspace{-0.15cm}
\begin{quote}
{\fontfamily{phv}\selectfont {\footnotesize 
	\underline{GET} \emph{advertiser.com?syncID=\textbf{user123}\&publisher=\textbf{website3.com}} \\
	\underline{Cookie}: \{cookie\_ID={\bf userABC}\}
	}}
\end{quote}
\vspace{-0.15cm}

\noindent
When advertiser.com receives the above request along with the cookie ID {\tt userABC}, it finds out that {\tt userABC} visited website3.com. To make matters worse, advertiser.com also learns that the user whom tracker.com knows as {\tt user123}, and the user {\tt userABC} is basically one and the same user.
Effectively, \csync\ enabled advertiser.com to collaborate with tracker.com, in order to: (i) find out which users visit website3.com, and (ii) synchronize (i.e., join) two different identities (cookies) of the same user on the web.

\subsection{\cs, Personalized Advertising \& Privacy Implications}
Digital advertising has moved towards a more personalized model, where ad-slots are purchased programmatically (e.g., Real Time Bidding (RTB) based auctions) in auctions, based on how well the profile of the visitor matches the advertised product.
Consequently, advertisers need to obtain user data (e.g., interests, behavioral patterns) to use as input in their sophisticated decision engines.
The core component for this data purchasing/sharing includes \csync~\cite{Ghosh:2015:MME:2764902.2745801}.
It allows trackers and advertisers to perform common user identification and by participating in data markets enrich their knowledge base with user information from several data sources.

There are several privacy implications for the online users who access websites planted with such sophisticated tracking technologies.
Using \csync, in practice, advertiser.com learns that: (i) what it knew as {\tt userABC} is also {\tt user123}, and (ii) this user has just visited website3.com.
This enables advertiser.com to track a user to a much larger number of websites than was initially thought.
Indeed, by collaborating with several trackers, advertiser.com is able to track users across a wide spectrum of websites, even if those websites do not have any collaboration with advertiser.com.
 
To make matters worse, the ill effects of \csync\ may reach way back in the past - even up to the time before the invention of \csync.
Assume, for example, that someone manages to get access to all data collected by tracker.com, and all the data collected by advertiser.com  (e.g., by acquisition~\cite{radioshack,facebookCS1}, merging or hacking of companies~\cite{facebookWhatsapp}).
In the absence of \csync, in those two datasets, our user has two different names: {\tt user123} and {\tt userABC}.
However, after one single \csync, those two different names can be joined into a single user profile, effectively merging all data in the two datasets.
Nowadays, such cases of \emph{server-to-server user data merges} are taking place at a massive scale~\cite{Englehardt:2016:OTM:2976749.2978313}, with the different web companies conducting mutual agreements for data exchanges or purchases, in order to enrich the quality and quantity of their user data warehouses~\cite{facebookCS1,dejavu}.

As if these threats to user privacy were not enough, \csync\ can rob users of the right to erase their cookies.
Indeed, when coupled with other tracking technologies (i.e., evercookie~\cite{zombieCookie}, or user fingerprinting~\cite{Eckersley:2010:UYW:1881151.1881152}), \csync\ may re-identify web users even after they delete their cookies.
Specifically, when a user erases her browser state and restarts browsing, trackers usually place and sync a new set of userIDs, and eventually reconstruct a new browsing history.
But if one of them manages to respawn~\cite{Acar:2014:WNF:2660267.2660347} its cookie (e.g. through evercookie~\cite{zombieCookie}), then through \csync, all of them can link the user's browsing histories from before and after her state erasure.
Consequently: (i) users are not able to abolish their assigned userIDs even after carefully erasing their set cookies, and (ii) trackers are enabled to link user's history across state resets.


\begin{figure}[t]
	\centering
	\includegraphics[width=0.9\columnwidth]{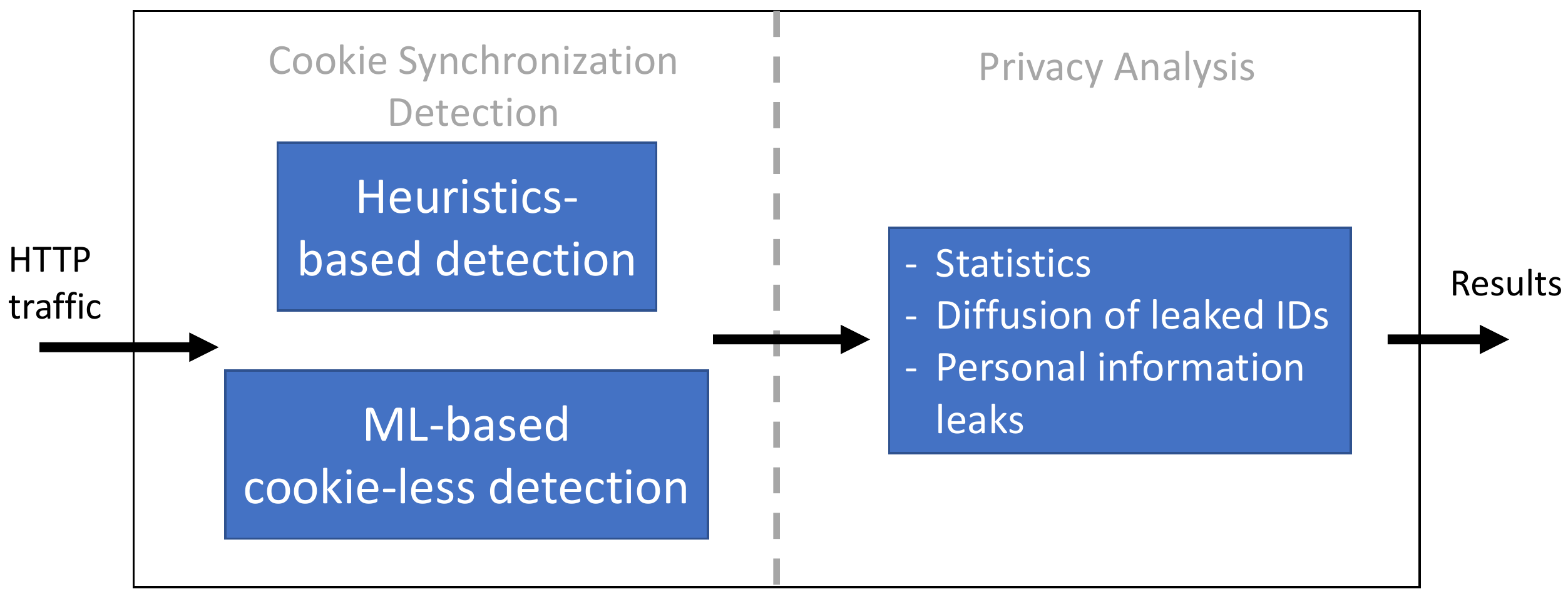}\vspace{-0.3cm}
	\caption{High-level overview of \toolname's internal components.}
	\label{fig:design}
\end{figure}
\vspace{-0.15cm}

\section{\cs\ Detection}
\label{sec:detection}

In this paper, we design \toolname, a holistic methodology to detect \csync\ events in real time, on the user side.
\toolname\ monitors the HTTP(S) traffic of the user on the browser level and detects userIDs when shared from one domain to the other.
To achieve that, it uses a (i) Heuristics-based (stateful) detection mechanism (Section~\ref{sec:heuristics}), where the IDs from cookies are tainted and alert is raised when they are exfiltrated to a domain other than the owner of the cookie.
However, as also presented in past studies~\cite{bashir2016tracing}, more and more companies include encryption (or cryptographic hashing) in the \csync-related APIs, thus concealing the synced IDs.
To deal with these cases, \toolname\ uses a (ii) ML-based (stateless) detection mechanism (Section~\ref{sec:id-less-detection}) capable of classifying with high accuracy such possible concealed synchronizations, without relying on any previously stored cookie IDs, but only using characteristics from the connections themselves.
Figure~\ref{fig:design}, provides a high level overview of \toolname's internal components.

For each \csync\ detection method, \toolname\ extracts its information flow: the chain of domains that share the synced ID, the domain that triggered the \csync\ event, the domains that (without having access to the website) used this sync request to set and sync their own userIDs (see Section~\ref{sec:trigger}). This way, our tool is able to measure the diffusion of anonymity loss for the given user by analysing what number of their overall userIDs budget got synced, and to how many 3rd-party domains.
In addition, by using a simple pattern matching technique, \toolname\ extracts possible personal information leaks tailored with the synced ID (see Section~\ref{sec:piileaks}).

Although there are several existing techniques for detecting ID-sharing events even when cookies are encrypted, accurate \csync\ detection in real time is a hard task.
The main advantages of our approach, contrary to existing detection mechanisms~\cite{Acar:2014:WNF:2660267.2660347,Englehardt:2016:OTM:2976749.2978313,bashir2016tracing} are as follows:
(i) It offers the ability to detect synchronizations when the userID is embedded not only in the URL's parameter, but also in its path (either in case of request/response URL or Location URL of the referrer).
(ii) By filtering-out domains of the same provider, our approach can discriminate between intentional \csync\ and unequivocally legitimate cases of internal ID sharing, thus avoiding false positives. 
(iii) It is capable of detecting \cs\ \emph{at real time}, even when shared IDs are encrypted.

\subsection{Heuristics-based detection}\label{sec:heuristics}

 \begin{table}[t]
 	\caption{Examples of userIDs synchronized among various domains.}\vspace{-0.4cm}
 	\label{tbl:examples}
	\centering
	{\footnotesize 
 		\begin{tabular}{l}
 			{\bf URLs of \cs\ HTTP Requests} \\
 			\toprule
 			{\bf 1.} a.atemda.com/id/csync?s=\textbf{\textcolor{heraldBlue}{L2zaWQvMS9lkLzMxOUwOTUw}} \\
 			{\bf 2.} bidtheater.com/UserMatch.ashx?bidderid=23\& \\
 			bidderuid=\textbf{\textcolor{heraldBlue}{L2zaWQvMS9lkLzMxOUwOTUw}}\&expiration=1426598931 \\
 			{\bf 3.} d.turn.com/r/id/\textbf{\textcolor{heraldBlue}{L2zaWQvMS9lkLzMxOUwOTUw}}/mpid/ \\
 			\bottomrule
 		\end{tabular}
}
\vspace{-0.15cm}
\end{table}

Technically, as we see in Table~\ref{tbl:examples}, \csync\ is nothing more than a request from the user's browser to a 3rd-party domain carrying (at least one) parameter that constitutes a unique ID set by the calling domain.
However, what \csync\ typically enables is a multiple, back-to-back operation with several 3rd-party domains getting updated with one particular ID.
This multiple synchronization happens by utilizing URLs of HTTP requests (i.e., the Location HTTP header), in which the cookie ID (i.e., the userID) of the triggering domain is embedded.
The userID may be embedded in the: (i) parameters of the URL, (ii) URL path, or (iii) referrer field.
In some cases, detection may be straightforward: one can simply look for specific parameter names (e.g., syncid, user\_id, uuid).
However, different companies use different APIs and parameter naming; relying only on string matching for \cs\ detection will lead to a large number of false negatives in case of newcomer syncing domains.
To remedy this, by extending previous work~\cite{lukasz2014selling-privacy-auction}, we design a stateful heuristics-based detection algorithm, which relies on the previously set cookies to taint userIDs that may get synced with domains different than the cookie setter\footnote{A known limitation of this approach is its inability of capturing a small portion of 1st-party cookies set by javascript\cite{jscookies}}.
In particular, our \cs\ detection methodology includes the following steps, which are also illustrated in Figure~\ref{fig:detection}:

\begin{enumerate} [leftmargin=1.5em,topsep=0pt,itemsep=0pt]
\item We extract all cookies set on the user's browser. To accomplish that, we parse all HTTP requests in our dataset and extract all \emph{Set-Cookie} requests.
	\begin{enumerate}
	\item We filter out all session cookies. These are cookies without expiration date, that get deleted after the end of a session.
	\item We parse the cookie value strings using common delimiters (i.e. ``:'', ``\&''). By extracting potentially identifying strings (cookie IDs), we create a list with the cookie IDs that could uniquely identify the user in the future.
	\end{enumerate}

\begin{figure}[t]
	\centering
	\includegraphics[width=0.8\columnwidth]{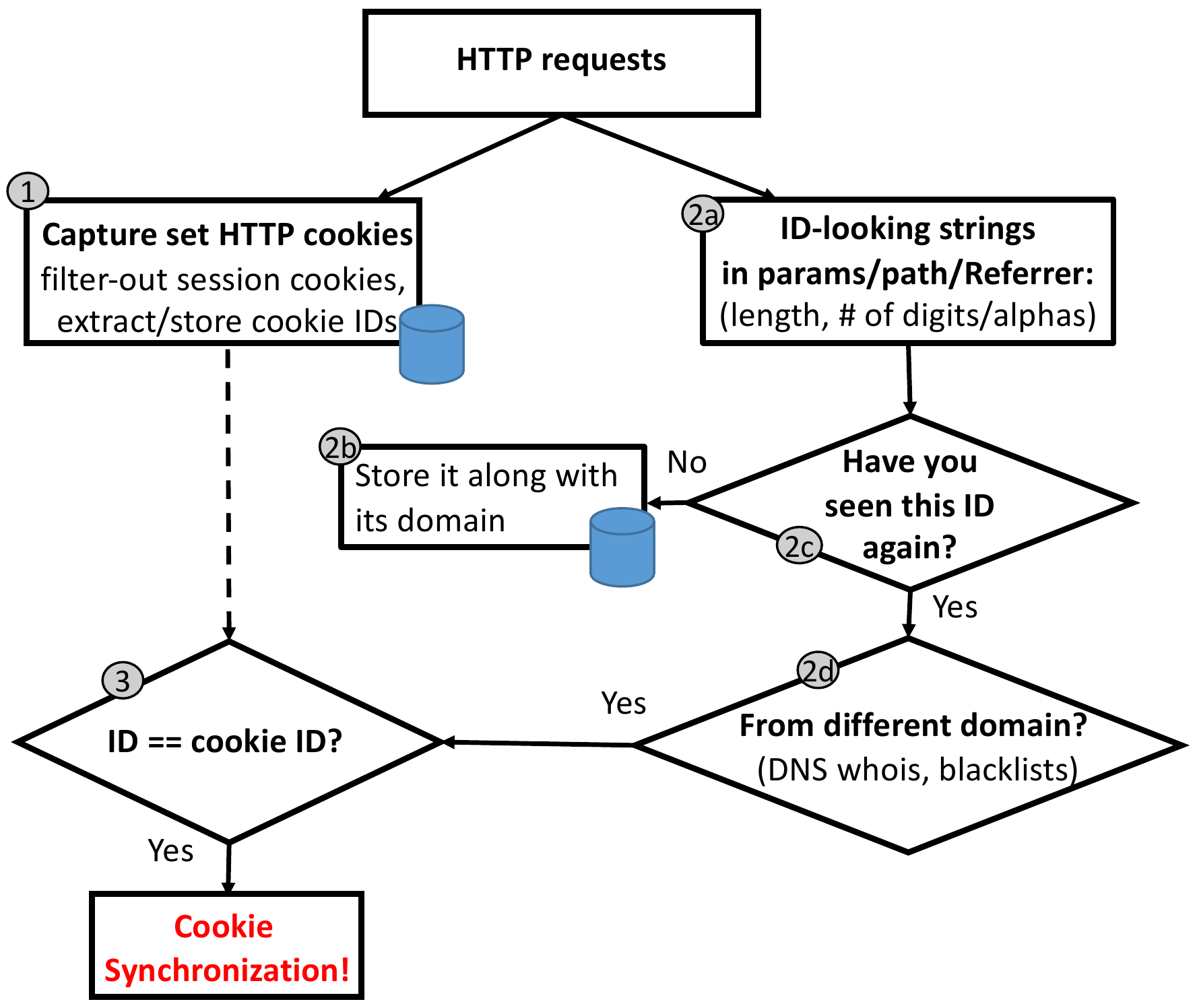}
	\caption{Heuristics-based \csync\ detection mechanism.}
	\label{fig:detection}
\vspace{-0.15cm}
\end{figure}

\item We detect possible ID-sharing events in the HTTP requests: 
	\begin{enumerate} [leftmargin=1.5em,topsep=0pt,itemsep=0pt]
	\item We identify ID-looking strings that are unique per user, carried either (i) as parameters in the URL, (ii) in URL path, or (iii) in the referrer field.
	As ID-looking strings, we define strings with specific length ($>10$ characters) -- false positives do not matter at this point.
	\item If this ID is seen for the first time, it is stored in a hashtable along with the URL's domain (receiver of the ID).
	\item If this ID has already been seen, we consider it as a shared ID and the requests carrying it as \emph{ID-sharing requests}.
	\item To check if the above ID-sharing requests regard different domains, we use several external sources (DNS whois, blacklists etc.) to filter-out cases where the IDs are shared among domains owned by the same provider (e.g., amazon.com and amazonaws.com)~\cite{mayer2011tracking}.
	This way, our approach can discriminate between intentional ID leaking and legitimate cases of internal ID-sharing, thus avoiding false positives.
	\end{enumerate}
\item Finally, to verify if the detected shared ID is a userID able to uniquely identify a user, we search this ID within the list of cookie IDs extracted in the first step.
If there is a match, then we consider this request as \csync.
\end{enumerate}

\subsection{Cookie-less detection}\label{sec:id-less-detection}
It is apparent that in order for the above methodology to be a viable \csync\ detection method, cookie IDs need to be shared in plaintext.
However, major web companies such as DoubleClick~\cite{googleCS} have started encrypting the cookie ID in an attempt to protect the actual cookie from being revealed to unwanted parties that may snoop the user's traffic (plugins or even ISPs).

While the former case is obvious why companies would want to block such snooping, the latter case may not be clear why and, thus, we discuss this case further.
In particular, under the traditional, plaintext case of cookie ID syncing, the same source company can sync independently with multiple 3rd-parties for the same user cookie ID.
Thus, no-one forbids these other 3rd-parties from syncing their IDs with each other, and find out that they have information about the same user, something that goes beyond what the source company intended to do (top example of Table~\ref{tbl:cs-thirdparties}).
With hashing or encryption of the cookie ID, these 3rd-parties are unable to do this syncing (bottom example of Table~\ref{tbl:cs-thirdparties}).
As a consequence of this encryption, \csync\ events can proceed undetected, if the previously used detection method is employed.

To address this scenario, in this paper, we propose a novel method for identifying \csync\ which is oblivious to the IDs shared.
This mechanism is able to identify with high accuracy \csync\ events in web traffic, even when the leaked IDs are protected and cannot be matched.
To build this mechanism, we employ machine learning methods, which we train on the ground truth datasets created with the previous, heuristic-based technique.
In particular, we analyze various features extracted from the web traffic due to \csync, and train a machine learning classifier to automatically classify a new HTTP connection as being a \csync\ event or not.
Here, we make the assumption that the various features used to characterize, and eventually detect, \csync\ with plaintext IDs, are equally used, and have the same distributions and variability as in the \csync\ with encrypted IDs.
We believe this is a reasonable assumption, since the companies employing encrypted IDs are not expected to change the rest of their mechanism which delivers these IDs and triggers \csync\ with their partners; these companies only want to obfuscate the IDs to avoid further, and unwanted, \csync.

\begin{table}[t]
	\caption{Examples of \cs\ between 3rd-parties with plaintext and encrypted cookie IDs.}\vspace{-0.3cm}
	\label{tbl:cs-thirdparties}
	\footnotesize
	\begin{tabular}{p{8.4cm}}
		{\bf ID syncs beyond the 1-1 of source domain (plaintext):}\\ 
		\toprule
		Domain syncs ID with Tracker1 ID1; $\rightarrow$ ID1=ID;\\
		Domain syncs ID with Tracker2 ID2; $\rightarrow$ ID2=ID;\\
		Tracker1 syncs ID1 with Tracker2 ID2; $\rightarrow$ ID2=ID1=ID;\\ \hline
		{\bf ID syncs beyond the 1-1 of source domain (encrypted):}\\
		\hline
		Domain syncs h(ID,A) with Tracker1 ID1; $\rightarrow$ ID1=h(ID,A);\\
		Domain syncs h(ID,B) with Tracker2 ID2; $\rightarrow$ ID2=h(ID,B);\\
		Tracker1 syncs ID1 with Tracker2 ID2; $\rightarrow$ h(ID,A)!=h(ID,B), i.e., ID1!=ID2;\\
		\bottomrule
	\end{tabular}
\vspace{-0.15cm}
\end{table}

\begin{figure*}[t]
	\centering
	\begin{minipage}{0.24\textwidth}
		\centering
		\includegraphics[width=1.1\columnwidth]{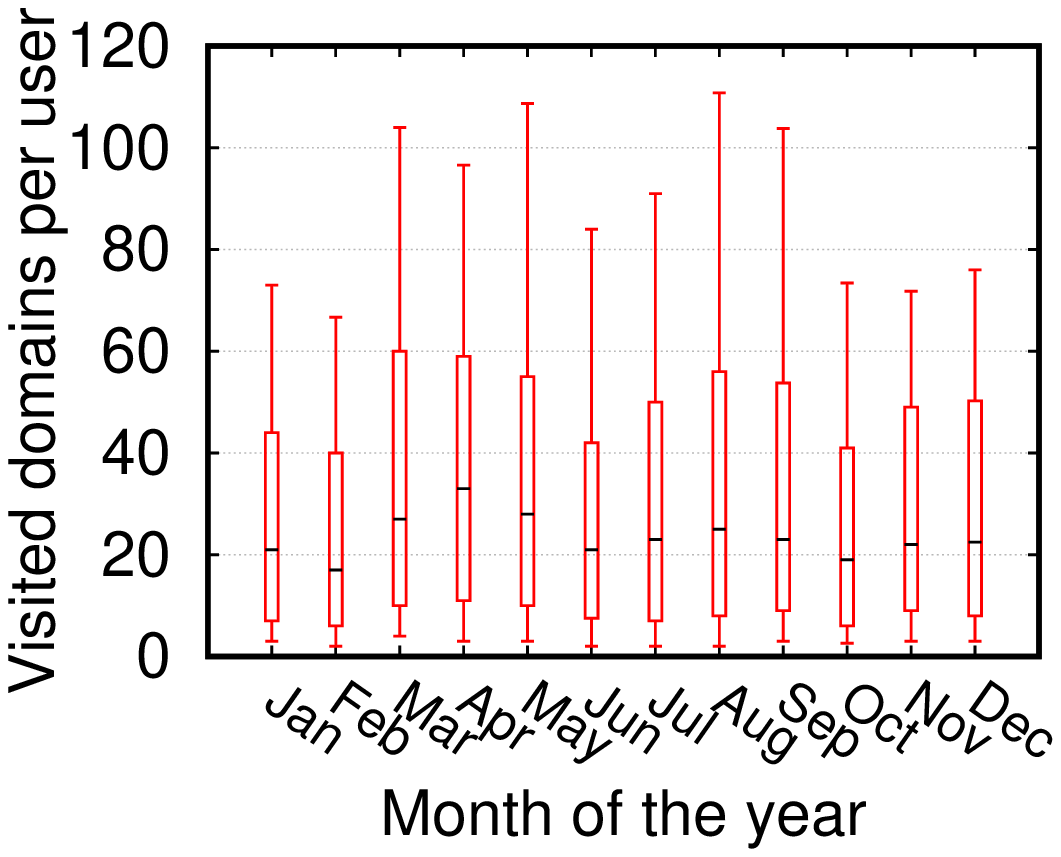}
		\caption{Distribution of number of unique domains visited per user, per month. The median user in our dataset visits 20 - 30 different domains per month.}
		\label{fig:domainsPerUser}
	\end{minipage}
	\hfill
	\begin{minipage}{0.24\textwidth}
		\centering
		\includegraphics[width=1.1\columnwidth]{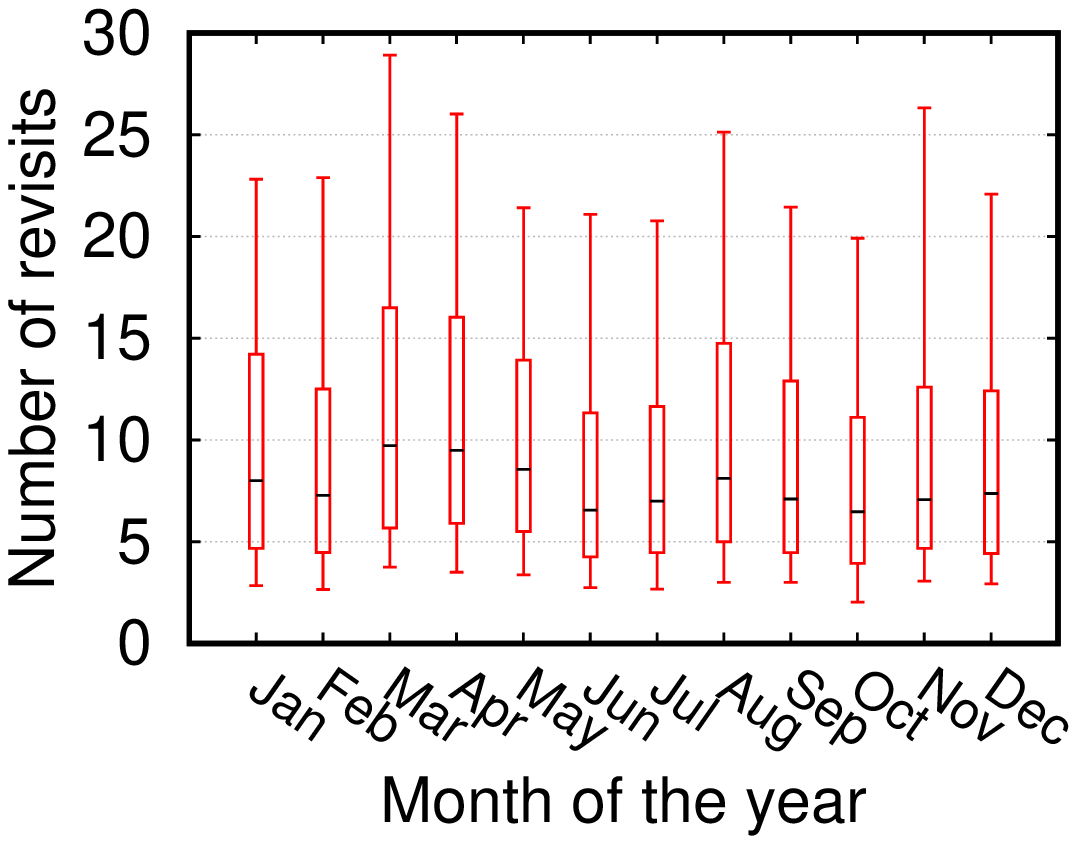}
		\caption{Distribution of number of times a user revisits the same domain per month. The median user revisits a domain around 7-10 times per month.}
		\label{fig:revisits}
	\end{minipage}
	\hfill
	\begin{minipage}{0.24\textwidth}
		\centering
		\includegraphics[width=1.1\columnwidth]{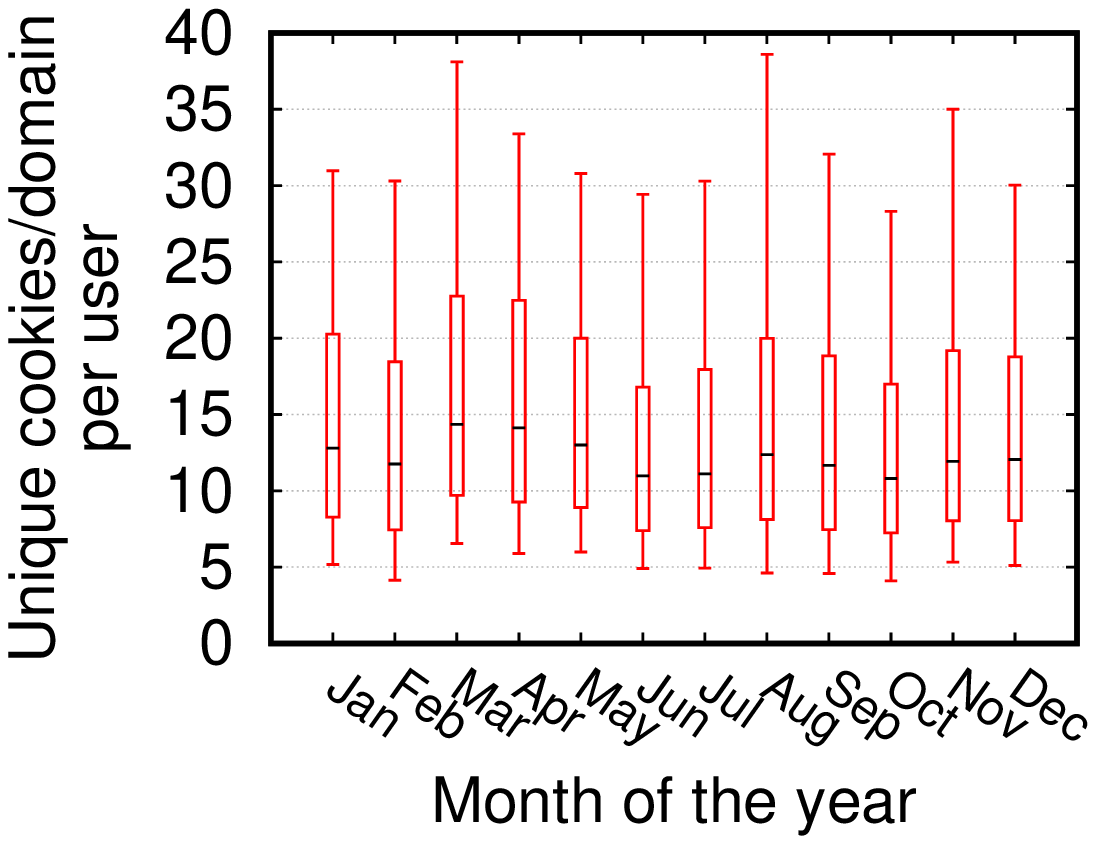}
		\caption{Number of (first and 3rd-party) cookies per domain per user. We see that the median user receives 12.25 cookies, on average, per visited website.}
		\label{fig:uniqCookieIDsperUser}
	\end{minipage}
	\hfill
	\begin{minipage}{0.24\textwidth}
		\centering
		\includegraphics[width=1.1\columnwidth]{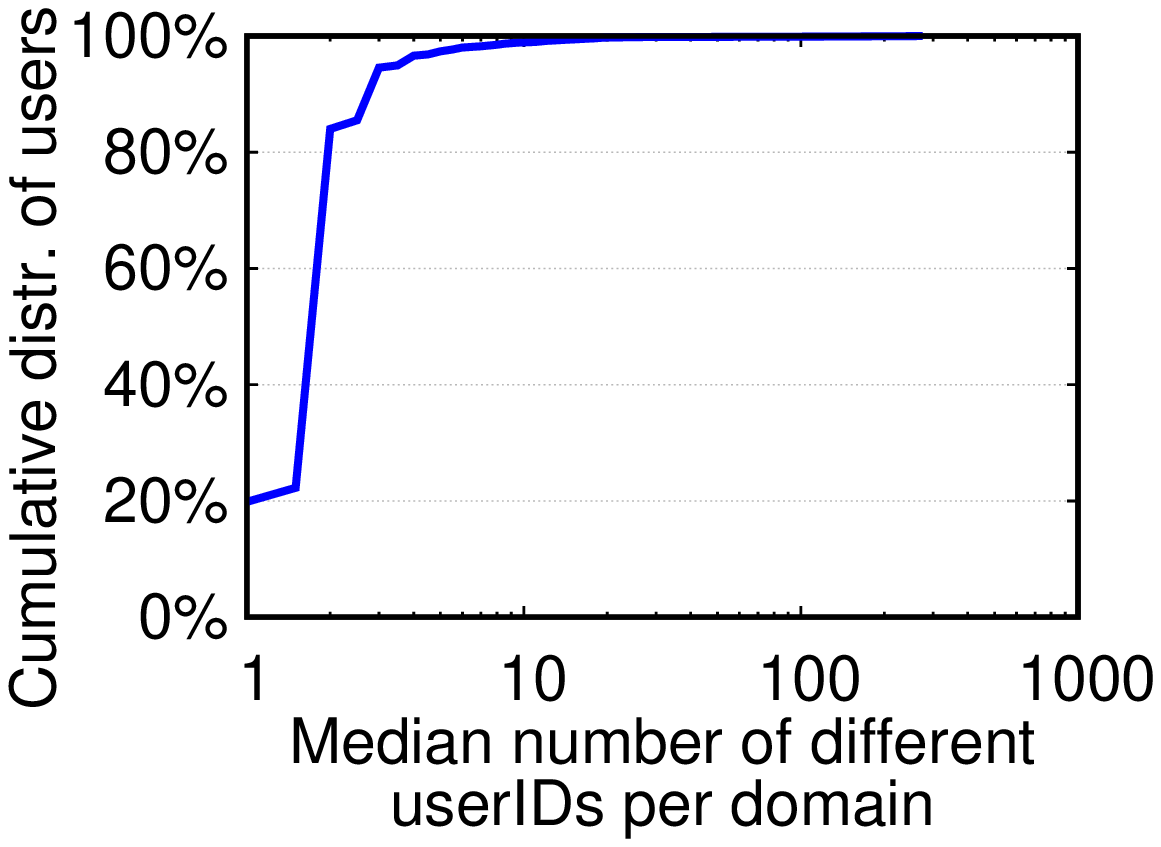}
		\caption{Unique userIDs set per domain, across the year. 80\% of users are known to a single domain with only $\sim$2 aliases, on average. }
		\label{fig:idsPerDomain}
	\end{minipage}
\end{figure*} 

For training the classifier, we extract various relevant features from the network traces.
As ground truth, we use confirmed \csync\ events which were detected with the heuristics-based method described earlier.
Beyond these confirmed events, there are \textit{id-sharing} events which were first selected by the method as potential \csync\ events, but eventually were rejected as \textit{non-\csync}, as they did not match cookie IDs already seen by the method (step 1 in Figure 3).

The features available for these network events can be several; we constrain the learning algorithm to use only features available at run time, and during the user's browsing to various websites:
\begin{itemize} [leftmargin=1.5em,topsep=0pt,itemsep=0pt]
\item EntityName: \{domain of recipient company\}
\item TypeOfEntity: \{Content, Social, Advertising, Analytics, Other\}
\item ParamName: \{aid, u, guidm, subuid, tuid, etc.\}
\item WhereFound: \{parameter in URL, parameter in Referrer, in the URL path\}
\item StatusCode: \{200, 201, 202, 204, etc.\}
\item Browser: \{Firefox, Chrome, Internet Explorer, etc.\}
\item NoOfParams: \{0, 1, 2, ..., etc.\}
\end{itemize}

Various machine learning algorithms can be applied: Random Forest, Support Vector Machines, Naive Bayes, and even more advanced methods such as Neural Networks.
However, a balance must be found between training the given algorithm to reach a good accuracy, the computation cost for this training, as well as the capability of the algorithm to be used at real time on the user device.

This method aims to address two possible scenarios that can arise while IDs are being shared:
First, we consider the realistic scenario that an already identified set of id-sharings are candidate \csync\ events (as found by the heuristics-based method), but cannot be validated as \csyncs\ because of the cookie ID being encrypted or unavailable, and therefore not matching the repository of IDs.
Second, we consider the scenario where various HTTP connections are ingested by the method, and it needs to decide at run-time which are \csync\ events and which are not.
This case is a more generalized version of the previous scenario, and attempts to detect \csync\ events, as an alternative method to the heuristic-based approach.
In either of the two cases, we follow a generally accepted methodology (e.g.,~\cite{papadopoulos2017imc-rtb-auctions}) that separates training such a machine learning model, from applying it at real-time.
The training can be performed offline, on an existing dataset (e.g., the one we collected, or from anonymous user network traffic donations), and the model trained can be distributed accordingly with the tool at hand.
Then, the tool can apply the classifier on each network connection under question, for real-time classification.

\section{Dataset}
\label{sec:Dataset}

\begin{table}[t]
	\caption{Summary of contents in our dataset.}\vspace{-0.4cm}
	\label{tbl:dataset}
	\centering
	\small
	\begin{tabular}{l|l||l|l}
		{\bf Description}	&	{\bf \#}		&	{\bf Description}		&	{\bf \#} \\ \hline
		Total mobile users	&	\totalUsers	&	Unique shared IDs ($S$)	&	68215 \\
		Requests captured	&	179M		&	Unique userIDs synced ($C$$\cap$$S$)	&	\csIDs \\
		Unique Cookies ($C$)	&	8.97M	&	\csync\ requests		&	\csNumber	\\
		ID sharing requests	&	412805		&						&				\\ \hline
	\end{tabular} 
\end{table}

In this section, we describe the data collection process and our year long dataset.
In order to collect data from real users, we set a group of proxies fronted by a load balancer, and gathered \totalUsers\ volunteering users residing in the same country.
These users agreed to strip their browsers from any previous state (i.e., cookies, cache, webStorage) and have their devices to continuously redirect their network traffic through our proxies for 12 consecutive months (2015-2016).
They signed a consent form allowing us to collect and analyze their data during this period, and publish any anonymized results.
They were well-aware of the purposes of the data collection, and were compensated with free data plan, as long as they were using the proxies.
Before the analysis was performed, all data were anonymized and never shared with any other domains.

Given the long duration of the experiment, and in order not to jeopardize the confidentiality of the volunteering users' secure sessions, we capture only their HTTP mobile traffic (same method can be applied for HTTPS).
On the server-side and based on the user agent of each request, we filtered out any possible app-related traffic. Overall, we collected a dataset containing a total of 179M requests, spanning an entire year.
Table~\ref{tbl:dataset} summarizes the contents and \csync\ findings in our dataset.

\begin{figure*}[t]
	\centering
	\begin{minipage}{0.24\textwidth}
		\centering
		\includegraphics[width=1.1\columnwidth]{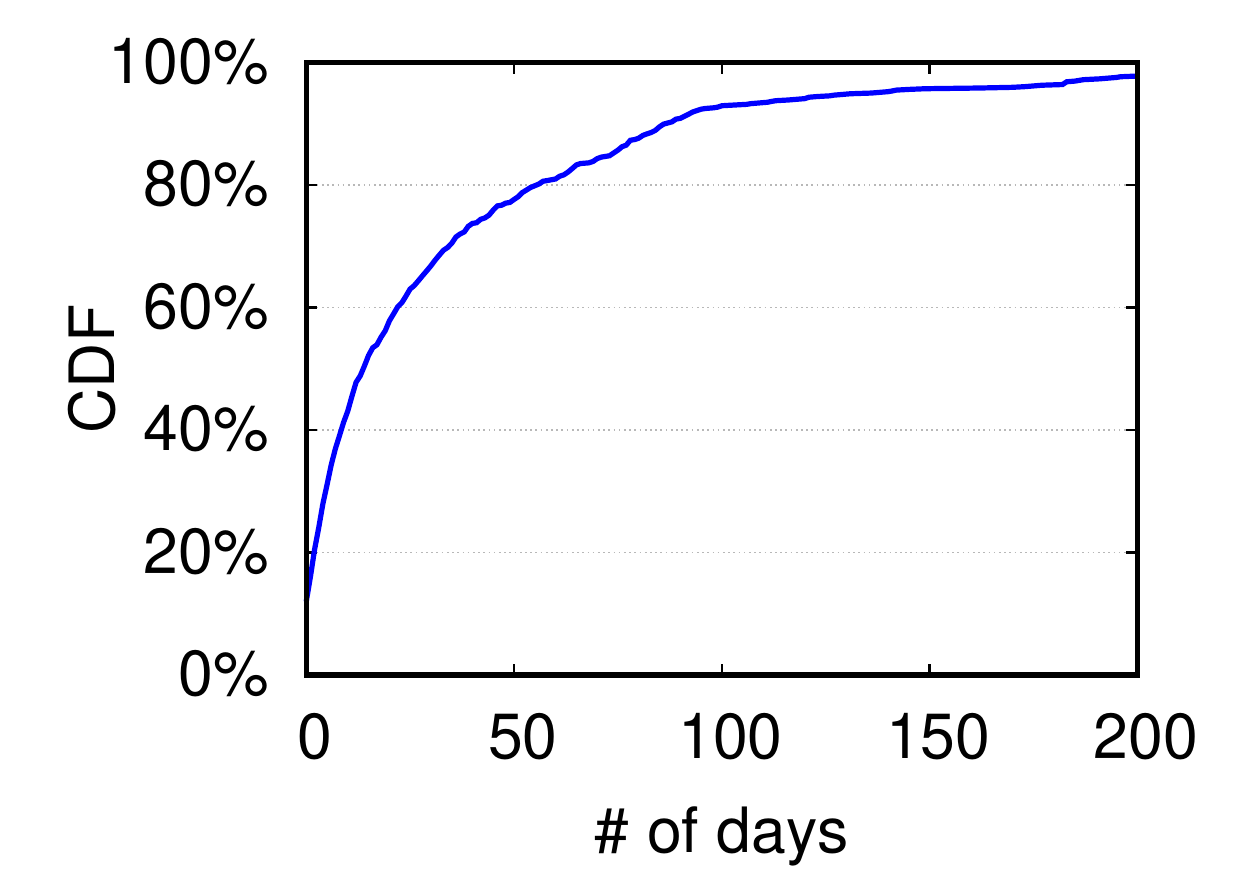}
		\caption{Distribution of time taken for first \csync\ to appear per user. 20\% of users get their first userID synced in 1 day or less.}
		\label{fig:userGotSyncedAfter}
	\end{minipage}
	\hfill
	\begin{minipage}{0.24\textwidth}
		\centering 
		\includegraphics[width=1.1\columnwidth]{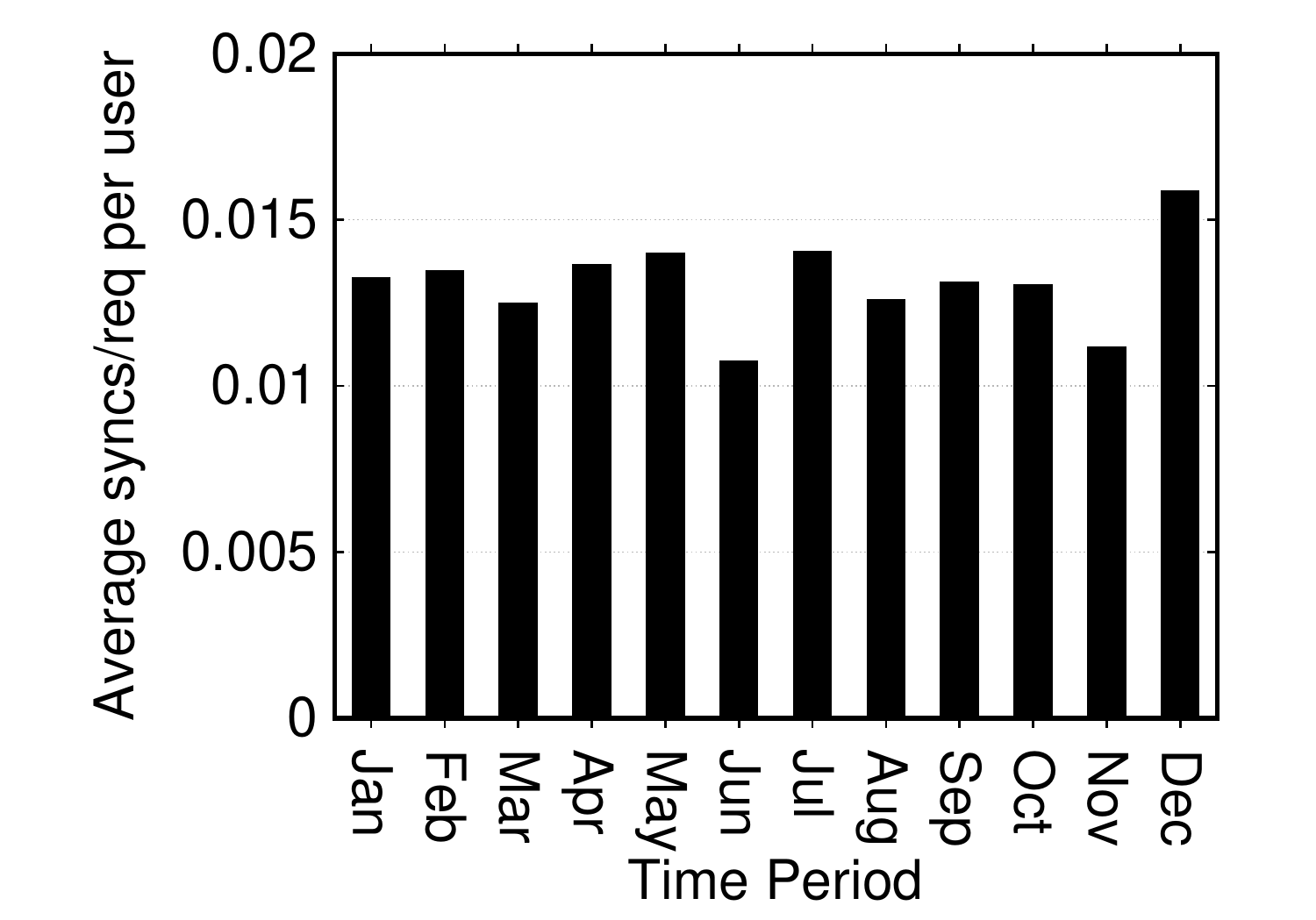}
		\caption{\csyncs\ per request for the average user across the year. The average user receives 1 synchronization every 68 requests.}
		\label{fig:csyncPerReq}
	\end{minipage}
	\hfill
	\begin{minipage}{0.24\textwidth}
	\centering
	\includegraphics[width=1.1\columnwidth]{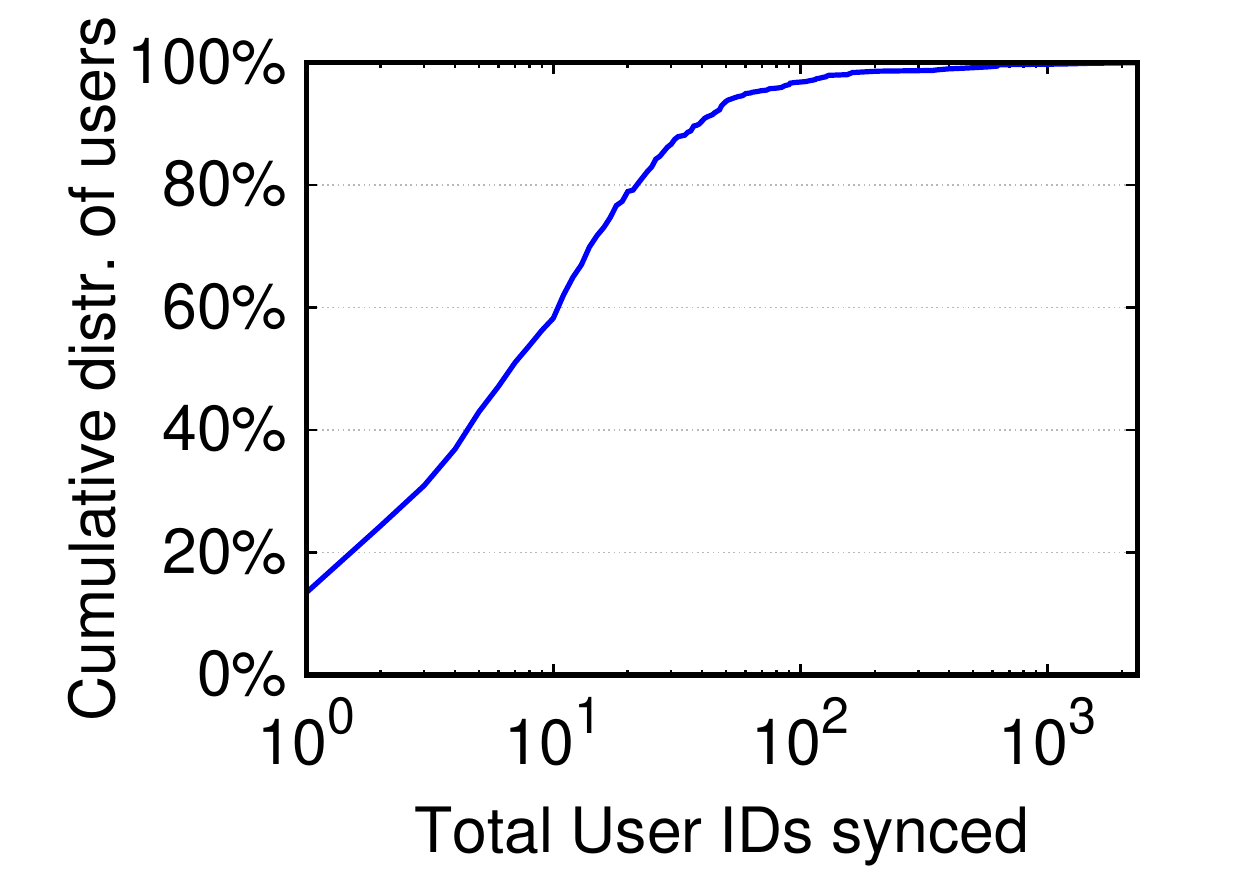}
	\caption{Distribution of the synced userIDs per user. The median user has 7 userIDs synced, when 3\% of users has up to 100.}
	\label{fig:csync_IDsPerUser}
\end{minipage}
\hfill
\begin{minipage}{0.24\textwidth}
	\centering
	\includegraphics[width=1.1\columnwidth]{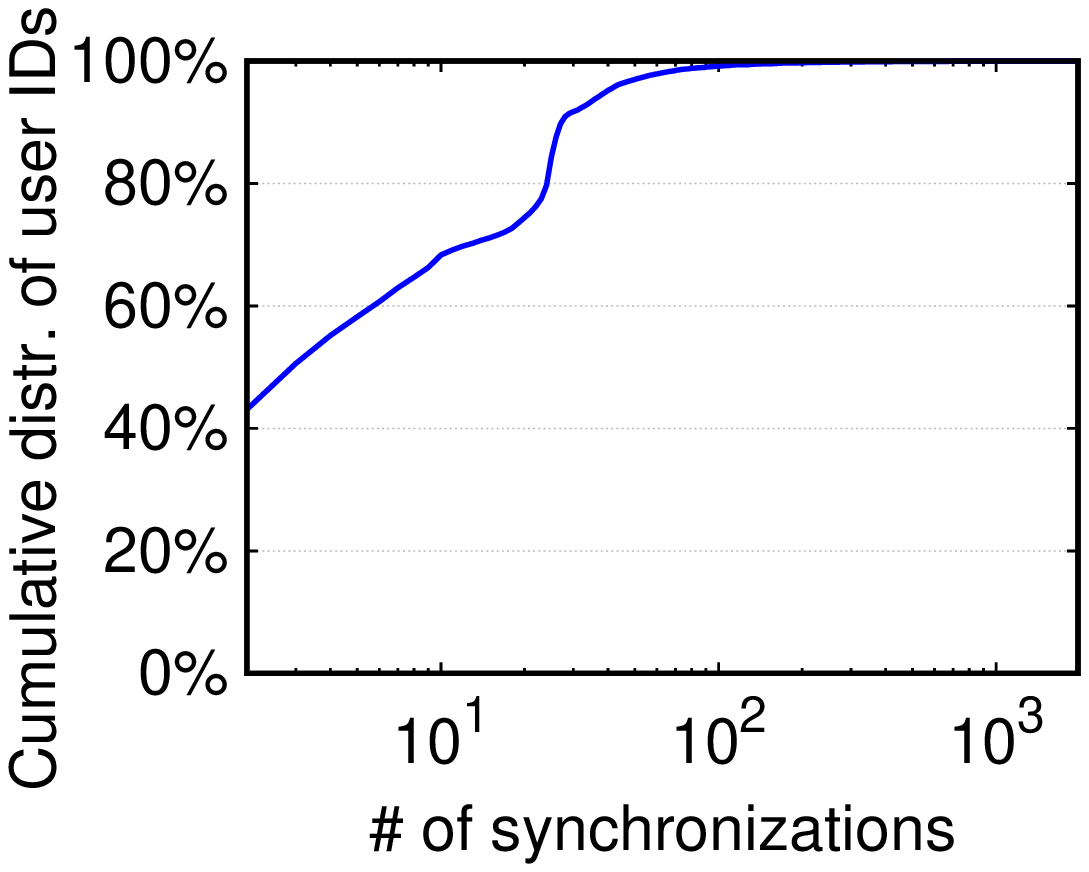}
	\caption{Distribution of synchronizations per userID. The median userID gets synced with 3.5 different domains.}
	\label{fig:csync_totalSyncsPerID}
\end{minipage}
\end{figure*}

\noindent{\bf Users:}
To analyse our dataset, we create a simple weblog parser.
As noted earlier, this dataset consists of web browser traffic from the mobile devices of \totalUsers\ users.
After separating the flows of each one of them, we produced their timelines, and in Figure~\ref{fig:domainsPerUser}, we present the number of different domains each user visits per month.
As we see from the distribution (Percentiles: 10th, 25th, 50th, 75th, 90th), the median user in our dataset visits 20-30 different domains depending on the month (we observe a seasonal phenomenon with increases during spring break and summer holidays).

Similarly, in Figure~\ref{fig:revisits}, we present the number of times each of these domains gets revisited by the median user in our dataset.
In every revisit, there is a new request that asks the user's browser if there is a previously set cookie. If this \emph{Get-Cookie} request regards a previously set cookie, this means that the domain already knows the user and we consider it as a revisit.
We observe that the median user revisits around 7-10 times the same domain from their mobile browser.
Again, we observe the seasonal phenomenon as earlier, when users tend to have more time to browse the web.
Also, the 75th percentile of the users may revisit the same website more than 15 times (March, August).

\noindent{\bf Cookies:}
To have a good view of the cookie activity of users, we extract all (1st \& 3rd-party) cookies set in users' browsers across the year, and in Figure~\ref{fig:uniqCookieIDsperUser}, we plot the distribution of number of cookies per visited website.
The median user receives a fairly constant number of 12.25 cookies per visited website per month, on average.

Next, we extract the unique identifiers (i.e., cookie ID) set in these cookies.
A cookie ID constitutes a unique string of characters that websites and servers associate with the browser and, thus, the user who stored the cookie.
Thereby, here, we consider a cookie ID as a unique user identifier called userID.
As it can be seen in Table~\ref{tbl:dataset}, in our dataset, there are almost 9 million such unique IDs.

In Figure~\ref{fig:idsPerDomain}, we plot the distribution of the number of unique userIDs assigned to the users per domain.
The vast majority of the users (80\%) receive, on average, only 2.2 userIDs per domain, across the year.
This means that users tend not to erase their cookies frequently, thus, allowing web domains to accurately identify them through time and during the users' browsing.
Only 1.13\% of users erase their cookies (either manually or by browsing with Private/Incognito Browsing), receiving more than 9.5 different userIDs per domain, on average.
This means that it is very rare for a domain to meet a previously known user with a different alias.

\section{Privacy Analysis}
\label{sec:measurements}

\begin{table}[t]
	\caption{Breakdown of the \csync\ triggering factors.} \vspace{-0.3cm}
	\label{tbl:initiator}
	\centering
	\small
	\begin{tabular}{lp{5cm}|l}
		&{\bf Initiator} & {\bf Portion} \\ \hline
		(i) & Publisher syncs its userID & 2.692\% \\ \hline
		(ii) & Embedded 3rd-party triggers syncing of its own set userID & 49.668\% \\ \hline
		(iii) & 3rd-party uses sync request to share its own set userID & 45.697\% \\ \hline
		(iv) & 3rd-party uses sync request to share with other domains the publisher's set userID & 0.2658\% \\ \bottomrule
	\end{tabular}
\vspace{-0.15cm}
\end{table}

By applying \toolname\  in our dataset, we find several IDs passed from one domain to another.
We detect 68215 such unique shared IDs.
From these IDs, \csIDs\ were actually cookie IDs from previously set cookies that were synced among different domains.
In total, these cookie IDs were found in \csNumber\ synchronization events (see Table~\ref{tbl:dataset} for a summary).
From the \csyncs\ detected in our dataset, userIDs were found in 91.996\% of the cases inside the URL parameters, 3.705\% in the Referrer URL, and in 3.771\% of the cases in the URL path.
Thus, \toolname\ was able to detect 3.771\% more cases of \csyncs\ than existing detection methods~\cite{lukasz2014selling-privacy-auction,Acar:2014:WNF:2660267.2660347,www18adcost}.

\subsection{Initiation of \cs}
\label{sec:trigger}
First, we correlate the cookie domain (the setter), the synchronizing request's Referrer field, and the publisher that the user visited, in order to extract the domain that triggered the \csync\ on the user's browser.
As seen in Table~\ref{tbl:initiator}, there are 4 distinct cases:
(i) the \csync\ was initiated by the publisher who syncs the userID he assigned for the user: we find 2.692\% of these cases in our dataset,
(ii) the synchronization was initiated from a publisher's iframe, by the guest 3rd-party which syncs its own userID (49.668\%),
(iii) a 3rd-party which participated in a previous synchronization (case (i) or (ii) above) and uses the sync request to share its own userID (45.697\%).
Lastly, there is a rare case (iv), where a 3rd-party participating in a previous synchronization of the publisher's userID (case (i)) initiates a new round of syncs while it continues to share the publisher's userID (0.2658\%).
Obviously, in case (iv), the initiating 3rd-party shares with its 3rd-party affiliates, a userID assigned by a domain (the publisher) beyond its control, and possibly awareness.\footnote{We reported all such cases and notified the respected publishers.}

\begin{figure*}[t]
	\centering
	\begin{minipage}{0.24\textwidth}
		\centering
		\includegraphics[width=1.1\columnwidth]{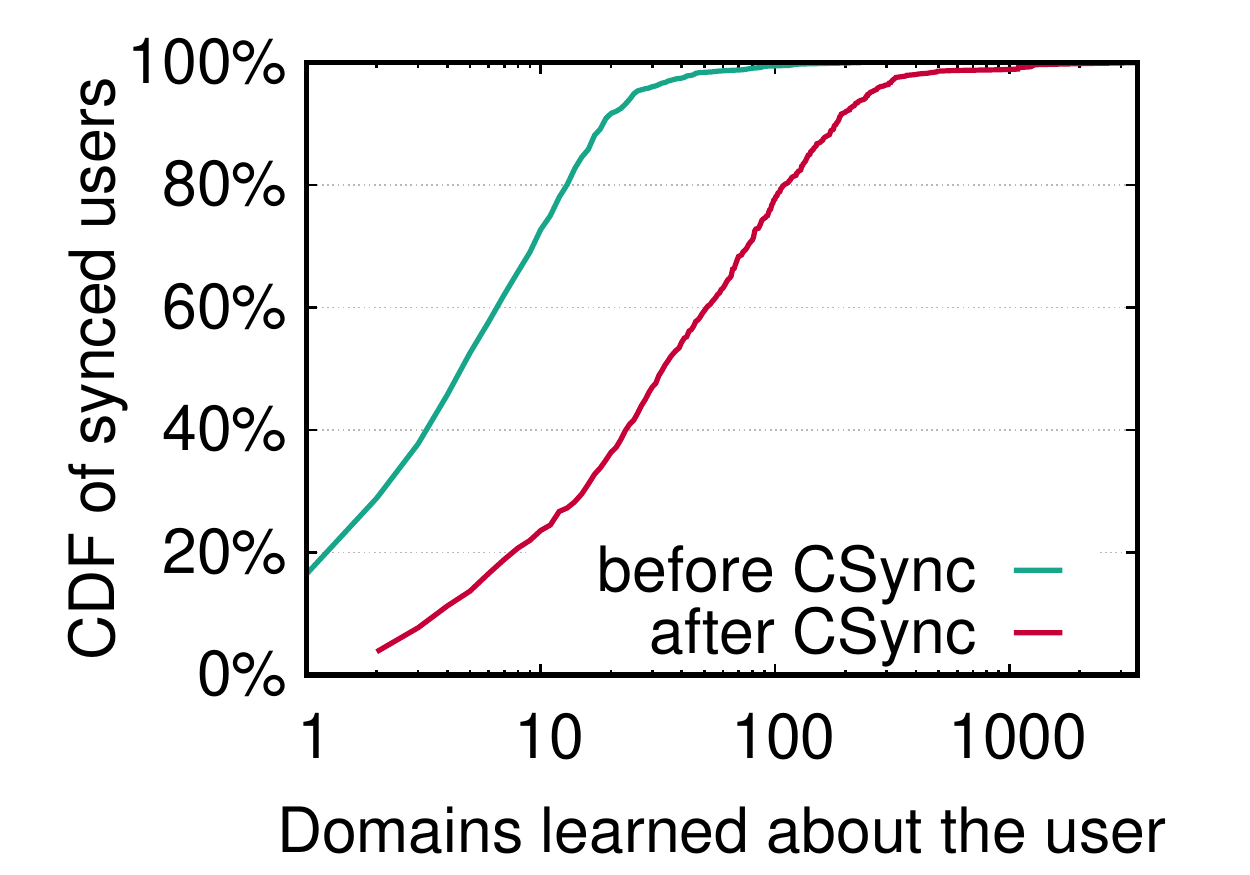}
		\caption{Distribution of domains learning at least a userID of a user (with/without the effect of \csync). After syncing, the domains that learned about the median user grew by 6.75$\times$.
		}
		\label{fig:IDlearners}
	\end{minipage}
	\hfill
	\begin{minipage}{0.24\textwidth}
		\centering
		\includegraphics[width=1.1\columnwidth]{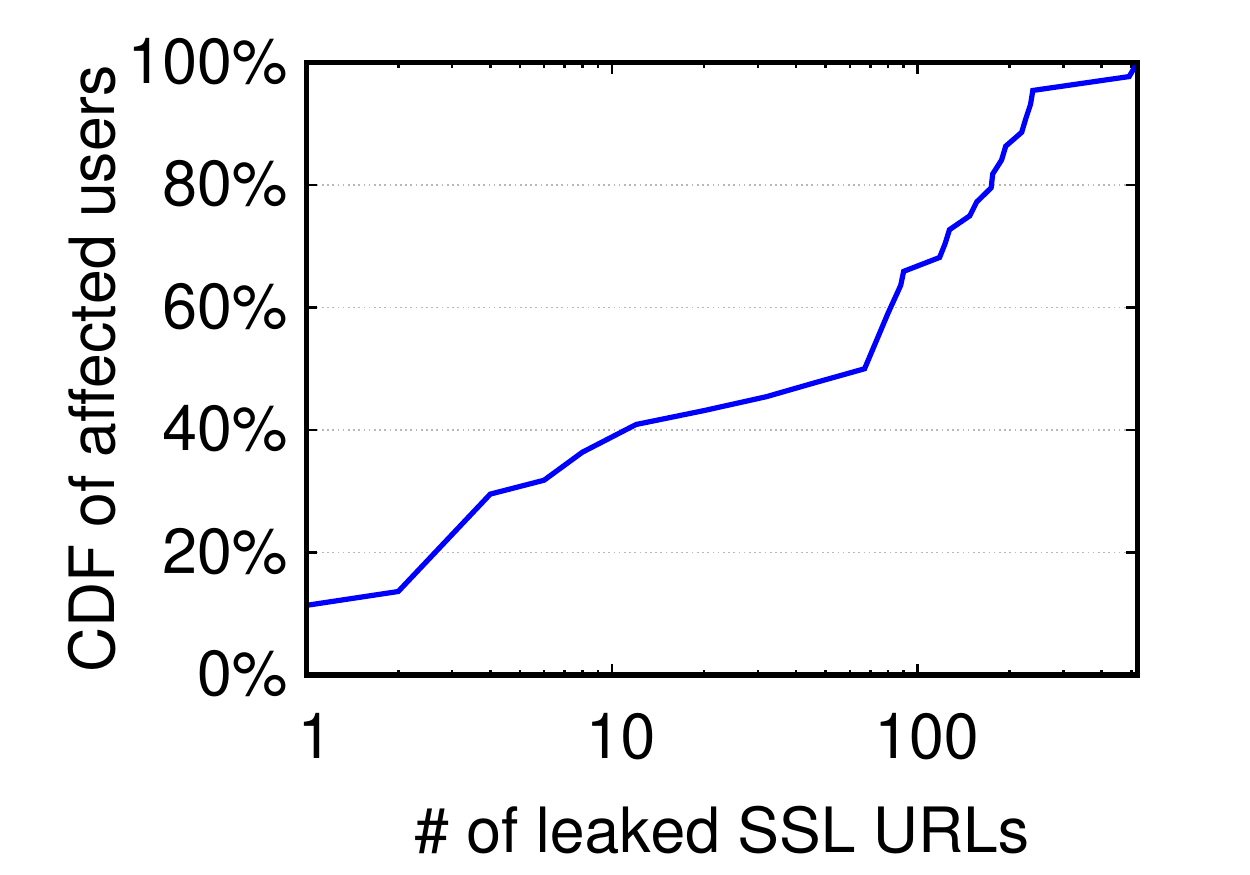}
		\caption{Distribution of the leaked TLS URLs per affected user. The median user has 70 TLS URLs leaked 
			through \cs, when the 90th percentile has up to 226 TLS URLs leaked.}
		\label{fig:sslLeak}
	\end{minipage}
	\hfill
	\begin{minipage}{0.24\textwidth}
		\centering\vspace{0.2cm}
		\includegraphics[width=1.05\columnwidth]{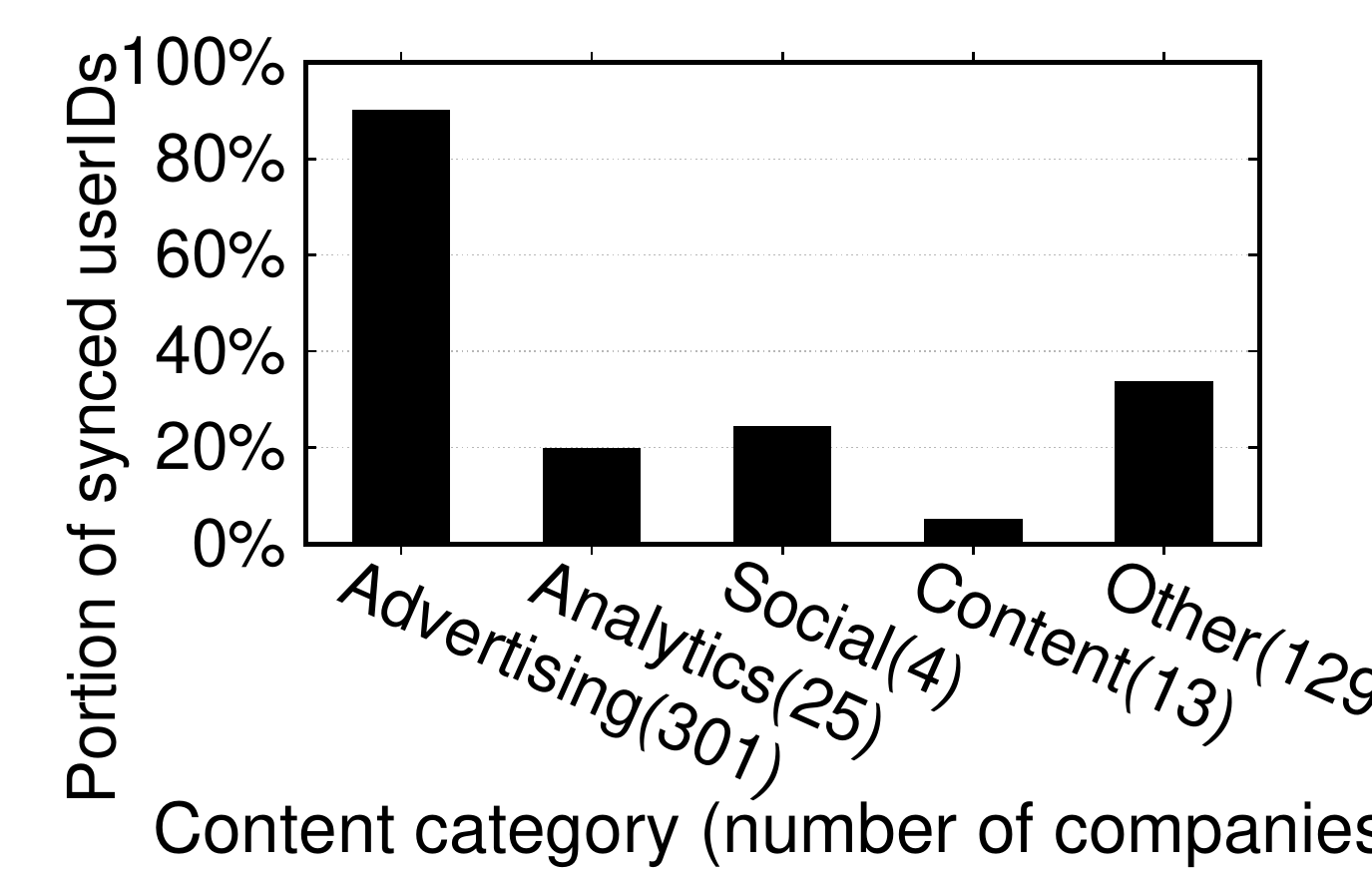}
		\vspace{-0.1cm}
		\caption{Portion of synced userIDs learned per content category. As expected, ad-related companies learned the vast majority (90\%) of the total synced userIDs in our dataset.}
		\label{fig:IDsPerCat}
	\end{minipage}
	\hfill
	\begin{minipage}{0.24\textwidth}
		\centering
		\vspace{0.3cm}
		\includegraphics[width=1.1\columnwidth]{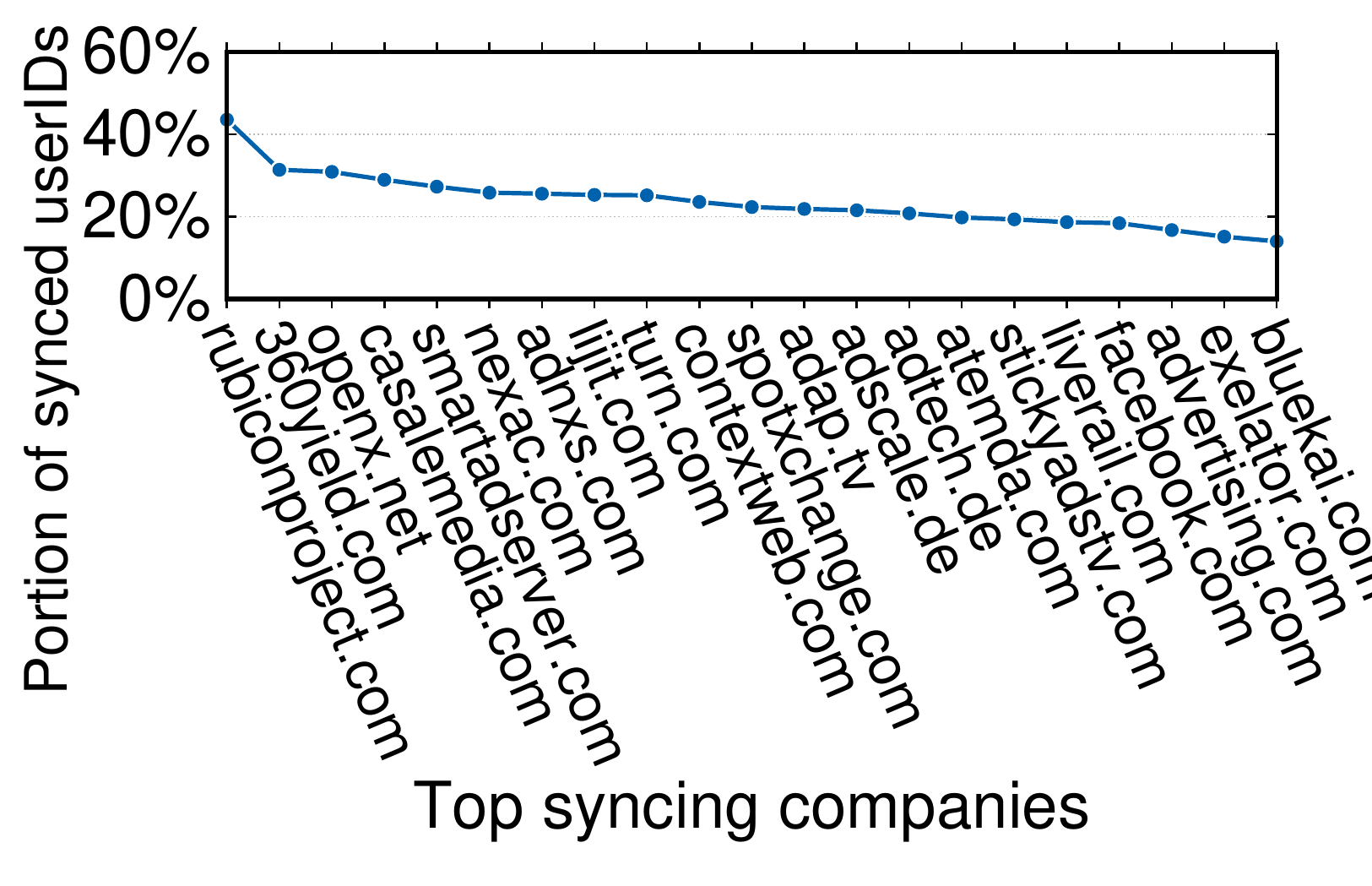}
		\vspace{-0.4cm}
		\caption{Portion of synced userIDs learned per tracker: 3 trackers learn more than 30\% of the total userIDs in our dataset; 14 trackers learn more than 20\% of userIDs each.}
		\label{fig:IDsPerComp}
	\end{minipage}
\end{figure*}

\subsection{How are users exposed to \csync?}
\csync\ impacts users' privacy by leaking assigned userIDs, and sharing them with 3rd-parties.
In our dataset, we see that for users with regular activity on the web ($>10$ requests per day), {\bf 97\% were exposed to \csync\ at least once}.
This means that \csync\ constitutes a phenomenon affecting the totality of online users.
Next, we study how long it takes for the first synchronization to happen, or in effect, how quickly a user gets exposed to \csync\ after she starts browsing.
Recall that as mentioned in Section~\ref{sec:Dataset}, all participating users, during bootstrapping phase, had all state from their browsers erased.
This means that our proxy was able to capture the very first cookie that was set during the user's monitoring period.
Of course, the time depends on the browsing patterns of each user, however, as we see in Figure~\ref{fig:userGotSyncedAfter}, a median user experiences at least one \csync\ within the first week of browsing.
In fact, a significant {\bf 20\% of users get their first userIDs synced in 1 day or less}.
It is worth noting that users tend to browse the same top websites repeatedly (e.g., facebook.com, twitter.com, cnn.com), so the set cookies are already shared and no sync is fired.

Next, we investigate if the synchronizations the users are exposed to, change over time.
Hence, we extract \csyncs\ per user, and normalize with the user's total number of requests. 
In Figure~\ref{fig:csyncPerReq}, we plot the average synchronizations per request across the year.
As shown, \csync\ is persistent through the duration of an entire year, with the user being exposed to a steady number of synchronizations across time.
Specifically, we see that {\bf the average user receives around 1 synchronization per 68 requests}.

Considering the different userIDs that trackers may assign to a user, in Figure~\ref{fig:csync_IDsPerUser}, we measure the number of unique userIDs that got synced per user.
Evidently, {\bf a median user gets up to 6.5 userIDs synced}, and 3\% of users has up to 100 userIDs synced.
It becomes apparent that the IDs of a user may leak to multiple 3rd-party domains through \csync.
To measure the userID leak diffusion, in Figure~\ref{fig:csync_totalSyncsPerID}, we plot the distribution of synchronizing requests per userID.
As we see, {\bf the median userID gets leaked, on average, to 3.5 different domains}.
There is also a significant 14\% that gets leaked to up to 28 different 3rd-parties.

To better understand the effect of \csync\ on the diffusion of the overall user privacy, we measure for each user the number of domains that learned about them (i.e., learned at least one of their userID) before and after \csyncs.
From the distribution in Figure~\ref{fig:IDlearners}, {\bf the domains that learned about the median user after \csyncs\ grew by a factor of 6.75}, and for 22\% of users this factor becomes $>10$.
This means that before the rise of \csync, when the user visited a website, the domains that could track them were only the publisher and the included 3rd-parties, but in an independent fashion.
However, with the introduction of \csync, the number of domains that can track the user drastically increased (6.75x for median user), severely decreasing their online anonymity.

 \begin{table}[b]
	\caption{Example of an \emph{ID Summary} stored on the user's browser. It includes userIDs and expiration 
		dates used for the particular user by 4 different domains.} \vspace{-0.4cm}
	\label{tbl:idsummary}
		\small
	\begin{tabular}{p{8.0cm}}
		{\bf ID Summary stored in cookie by adap.tv} \\ 
		\toprule
		``key=\textcolor{heraldBlue}{\bf valueclickinc}:value=708b532c-5128-4b00-a4f2-2b1fac03de81:expiresat=wed apr 01 15:03:42 pdt 2015,key=\textcolor{heraldBlue}{\bf mediamathinc}:value=60e05435-9357-4b00-8135-273a46820ef2:expiresat=thu mar 19 01:09:47 pst 2015,key=\textcolor{heraldBlue}{\bf turn}:value=2684830505759170345:expiresat=fri mar 06 16:43:34 pst 2015,key=\textcolor{heraldBlue}{\bf rocketfuelinc}:value=639511\\149771413484:expiresat=sun mar 29 15:43:36 pst 2015''\\\bottomrule
	\end{tabular}
\end{table}

\subsection{Buy 1 - Get 4 for free: ID bundling and Universal IDs}

We find {\bf 63 cases of domains which set on the users' browsers cookies with userIDs previously set by other domains}.
For example, we see the popular {\tt baidu.com}, the world's eighth-largest Internet company by revenue, storing a cookie with an ID $baiduid=\{idA\}$, and more than 5 different domains after this incident setting their own cookie using the same ID $baiduid=\{idA\}$.
This by-product of \csync, enables trackers to use \emph{universal IDs}, thus, bluntly violating the same-origin policy and merging directly (without background matching) the data they own about particular users.

In addition, we find {\bf 131 cases of domains storing in cookies the results of their \csyncs, thus composing \emph{ID Summaries}}.
In these summaries, we see the userIDs that other domains use for the particular user previously obtained by \csyncs.
An example of such summaries in JSON is shown in Table~\ref{tbl:idsummary}.
As one can see, the cookie set by {\tt adap.tv} includes the userIDs and cookie expiration dates of {\tt valueclick.com}, {\tt mediamath.com}, {\tt turn.com} and {\tt rocketfuel.com}.
In our dataset, we find at least 3 such companies providing \emph{ID Summaries} to other collaborating domains.
This user-side info allows (i) the synchronizing domains to learn more userIDs through a single synchronization request, and (ii) {\tt adap.tv} to re-spawn any deleted or expired cookies of the participating domains at any time, just by launching another \csync.

\subsection{Spilling userIDs out of TLS}
It is well-known that mixing encrypted and non encrypted sessions in TLS is a bad tactic~\cite{HTTPSmix}. 
Using TLS, everything except IPs and ports is encrypted~\cite{UrlInSSL}  in a HTTP connection.
As a result, although there are sophisticated estimation techniques~\cite{Gonzalez:2016:UPT:2987443.2987451}, no observer can monitor what the user is actually browsing over TLS, or the IDs assigned to her.
Given that our proxy is monitoring only HTTP traffic (see Section~\ref{sec:Dataset}), one would expect that no information from secure TLS sessions would be captured.
To our surprise, and as already found in~\cite{Papadopoulos:2018:ECM:3193111.3193117}, in our dataset we see userIDs 
that originated from TLS sessions getting leaked over plain HTTP to plain 3rd parties.
As a result, \emph{any curious, in-path observer (e.g., a snooping ISP) can eavesdrop the leaked userIDs}\footnote{As soon as we verified this leak, we notified publishers and also our volunteering users who updated their signed consent.}.
To make matters worse, we see that the same ID leaking requests have referrer fields\footnote{There are specific directives~\cite{rfc2616} for Referrer field hiding when referring over TLS visited domains.}, which leak the particular webpage the user visited over TLS (e.g., the particular article of the news site she read), thus leaking her interests.

To reproduce this leak, we manually visit the TLS protected websites where ID-spilling was found and we see it is caused by \csync\ events that sync a userID from a TLS cookie with non-TLS 3rd parties.
In Table~\ref{tbl:exampleAtt}, we present one such real case we observe.
As shown, while visiting over TLS the page \emph{https://financialexpress.com}, two \csyncs\ are performed: \emph{https://tapad.com} advertiser shares with \emph{http://switchadhub.com} and \emph{http://bluekai.com} the ID it assigned to the user.
This way, the latter two tracking domains sync their set-cookies with the one of \emph{https://tapad.com}.
However, by doing that over plain HTTP, the visited webpage gets leaked through the referrer field to a monitoring entity, even when users browse through proxies or VPN, or even Tor.
In addition, this entity from now on can re-identify the user in the web, just by monitoring the userIDs of cookies in requests destined to \emph{http://switchadhub.com} and \emph{http://bluekai.com}, even if the user's IP address is frequently changed.
Obviously, the more 3rd-parties were participating in the \cs, the easier it would be for the monitoring entity to capture HTTP requests loaded with these synced cookies and, thus, re-identify the user using a secure VPN.

\begin{table}[t]
	\caption{Example of ID-spill from SSL in our dataset.} \vspace{-0.4cm}
	\label{tbl:exampleAtt}
	{\footnotesize 
		\begin{tabular}{p{1.7cm}|p{6cm}}
			\toprule
			{\bf Role} & {\bf Domain}\\ \hline
			Visited website:& https://financialexpress.com\\ \hline
			Cookie setter: & https://tapad.com\\
			SetCookie:& \textbf{\textcolor{heraldBlue}{D0821FA0-8A80-4D9E-BC85-C40EAC4E4FF5}}\\ \hline
			Cookie syncer: & http://delivery.swid.switchadhub.com/adserver/user\_sync.php? SWID=cf43265166a9ccf5f6fd0472f23776fa\&sKey=PM2\& sVal=\textbf{\textcolor{heraldBlue}{D0821FA0-8A80-4D9E-BC85-C40EAC4E4FF5}}\\
			& \underline{referrer:} {\bf financialexpress.com} \\
			& \underline{Get-cookie:} \{cf43265166a9ccf5f6fd0472f23776fa\} \\ \hline
			Cookie syncer: &http://tags.bluekai.com/site/3096?id=\textbf{\textcolor{heraldBlue}{D0821FA0-8A80-4D9E-BC85-C40EAC4E4FF5}}\\
			& \underline{referrer:} {\bf financialexpress.com}\\
			&  \underline{Get-cookie:}  \{c57b29d1-f8e2-11e7-ac1b-0242ac110005\}\\ \bottomrule
		\end{tabular}
	}
\vspace{-0.25cm}
\end{table}

We measured such cases in our dataset and found {\bf 44 users (5\%) affected by the ID-spilling of \csync}.
The majority of leaked domains regard popular content providers, where an eavesdropper from the referrer field, apart from the domain, can also see sensitive information.
From Figure~\ref{fig:sslLeak}, the median (90th percentile) user has 70 (226) TLS URLs leaked through \csync.

\vspace{-0.2cm}
\subsection{Sensitive information leaked with userIDs}
\label{sec:piileaks}
The websites a user browses can easily leak through the referrer field during a \csync.
Moving beyond this type of basic leak, in our dataset we find several cases of privacy-sensitive information passed to the syncing domain regarding the particular user with the particular synced ID.
By deploying a simple string matching script, we look for keywords (e.g., gender, age, name, etc.) and find:
\begin{itemize} [leftmargin=1.5em,topsep=0pt,itemsep=0pt]
\item 13 syncs leaking the user's city level location
\item 2 syncs leaking the user's registered phone number
\item 10 syncs leaking the user's gender
\item 9 syncs leaking the exact user's age
\item 3 syncs leaking the user's full birth date
\item 2 syncs leaking the user's first and last name
\item 16 syncs leaking the user's email address
\item 4 syncs leaking user login credentials: username/password
\end{itemize}
\noindent
The above information constitutes not only a severe privacy threat for the user, but can also enables potential impersonation attacks.

\subsection{Who are the dominant \csync\ players?}
In order to assess the content that \csync\ parties provide, we extract all domains involved in \csync, and using EasyList, EasyPrivacy~\cite{easylist} and the blacklist of the Disconnect browser extension~\cite{disconnect} (enriched with our additions after manual inspection), we categorize them according to the content delivered.
This way, we create five categories of domains related with: (i) Advertising, (ii) Analytics, (iii) Social, (iv) 3rd-party content (e.g., CDNs, widgets, etc.), and (v) Other.

As we saw in Figure~\ref{fig:csync_totalSyncsPerID}, the median userID gets shared with more than one domains. 
We find that {\bf ad-related domains participate in more than 75\% of the overall \csyncs\ through the year}.
Consequently, as we can also observe in Figure~\ref{fig:IDsPerCat}, {\bf ad-related domains have learned as much as 90\% of all userIDs that got synced}, with Social and Analytics -related domains following with 24\% and 20\% respectively.
In Figure~\ref{fig:IDsPerComp}, we plot the top 20 companies\footnote{In a dataset with both HTTP and HTTPS traffic, market shares may differ since there are companies operating over SSL only (e.g., DoubleClick)} that learned the biggest portion of the total userIDs through \csyncs\ in our dataset.
Interestingly, the top 3 companies (i.e., rubiconproject.com, 360yield.com and openx.net) learned more than 30\% of all userIDs in our dataset each.
There is also a significant number (14) of companies that learn more than 20\% of userIDs.


\vspace{-0.15cm}
\section{Cookie-less Detection}
\label{sec:cookie-less-detection}

In this section, we explore two different scenarios outlined in Section~\ref{sec:id-less-detection} regarding the detection of \csync\ via ID sharing, while such IDs may be obfuscated to remove the possibility of matching them with past IDs shared between entities.
First, we explore the scenario where IDs have been shared, detected by the heuristic-based approach, but have not yet been confirmed as \csync\ events.
That is, we consider an already identified set of id-sharings, which are candidate \csync\ events, but cannot be validated as \csyncs\ because of the cookie ID being encrypted or unavailable (Section~\ref{sec:id-sharing-vs-cs}).
Second, we take a step back and consider the more general case where various HTTP connections are ingested by the method, and it needs to decide at run-time which are \csync\ events and which are not based on given features (Section~\ref{sec:random-vs-cs}).

Towards this end, we train and test the classifier in these two experiments.
We remind the reader of the assumption made earlier: the distributions of the features describing the \csync\ events with unencrypted IDs, have the same variability in the cases of encrypted IDs, and therefore can be used for the detection of such cases.
This assumption allows us to handle the problem as an out-of-sample estimation, leaving as future work the final validation with a set of ground-truth data of encrypted IDs that we also know their unencrypted versions.

\begin{table}[t]
	\caption{Performance of decision tree model trained on different subsets of features available at runtime for classification, given already identified id-sharing entries, and 10 cross-fold validation.
	}\vspace{-0.4cm}
	\label{tbl:ml-classifier-idsharing}
	\footnotesize
	\begin{center}
		\begin{tabular}{p{1.8cm}|p{0.1cm}|p{0.7cm}|p{0.7cm}|p{0.7cm}|p{0.7cm}|p{0.7cm}|p{0.5cm}}
			\hline
			Feature subset	&	F		&	TPR		&	FPR		&	PR		&	RC		&	FM			&	AUC	\\ \hline
			NoOfParams*	&	1		&	0.639	&	0.639	&	0.408	&	0.639	&	0.498		&	0.500	\\
			WhereFound+	&	1		&	0.643	&	0.610	&	0.612	&	0.643	&	0.535		&	0.602	\\
			StatusCode*+	&	1		&	0.648	&	0.619	&	0.723	&	0.648	&	0.523		&	0.633	\\
			TypeOfEntity*+	&	1		&	0.735	&	0.432	&	0.752	&	0.735	&	0.701		&	0.661	\\
			Browser*+		&	1		&	0.700	&	0.492	&	0.710	&	0.700	&	0.651		&	0.628	\\
			ParamName+	&	1		&	0.815	&	0.295	&	0.828	&	0.815	&	0.803		&	0.834	\\
			EntityName*+	&	1		&	0.803	&	0.295	&	0.806	&	0.803	&	0.793		&	0.854	\\  \hline
			\{id-less\}*		&	5		&	0.840	&	0.242	&	0.845	&	0.840	&	0.834		&	0.887	\\
			\{high imp.\}+	&	6		&	0.870	&	0.206	&	0.877	&	0.870	&	0.865		&	0.919	\\
			ALL			&	9		&	0.900	&	0.144	&	0.901	&	0.900	&	0.898		&	0.946	\\  \hline
		\end{tabular}
	\end{center}
\end{table}

\noindent\textbf{Data and Features: }
Based on the ground truth data presented earlier with the heuristic-based technique, we have $412.8k$ \textit{id-sharing} events, from which $263.6k$ are confirmed \csync, and $149.2k$ are identified as \textit{non-\csync}.
The features available for these events can be various, as already explained in~\ref{sec:id-less-detection}.
The ones we use are features available at run time, and during the user's browsing to websites.

\noindent\textbf{Algorithms: }
The final machine learning classifier used is decision tree-based.
Others like Random Forest, Support Vector Machines and Naive Bayes were tested, but the decision tree algorithm outperformed them, with significantly less computation and memory overhead.
Indeed, more advanced methods can be used, such as Neural Networks, but since the decision tree-based algorithm works very well, we leave this exploration for the future.

\begin{table}[tp]
	\begin{center}
		\caption{Performance of decision tree model trained on different subsets of features available at runtime for classification, given a pre-filter for ID-looking strings. All results besides the last row are with balanced dataset across the three classes, and 10-cross fold validation. The last row's results are computed given an unseen, and unbalanced test set, maintaining the original ratio of classes.
		}\vspace{-0.4cm}
		\footnotesize
		\label{tbl:ml-classifier-random}
		\begin{tabular}{p{1.8cm}|p{0.1cm}|p{0.7cm}|p{0.7cm}|p{0.7cm}|p{0.7cm}|p{0.7cm}|p{0.5cm}}
			\hline
			Feature subset	&	F		&	TPR		&	FPR		&	PR		&	RC		&	FM			&	AUC	\\ \hline
			NoOfParams*	&	1		&	0.541	&	0.314	&	0.584	&	0.541	&	0.495		&	0.706	\\
			StatusCode*+	&	1		&	0.666	&	0.229	&	0.673	&	0.666	&	0.598		&	0.764	\\
			TypeOfEntity*+	&	1		&	0.760	&	0.162	&	0.724	&	0.760	&	0.695		&	0.834	\\
			EntityName*+	&	1		&	0.865	&	0.075	&	0.863	&	0.865	&	0.860		&	0.962	\\
			ParamName+	&	1		&	0.870	&	0.083	&	0.878	&	0.870	&	0.859		&	0.953	\\  \hline
			\{id-less\}*		&	4		&	0.904	&	0.057	&	0.904	&	0.904	&	0.898		&	0.973	\\
			\{high imp.\}+	&	4		&	0.919	&	0.051	&	0.923	&	0.919	&	0.914		&	0.978	\\
			ALL			&	5		&	0.920	&	0.051	&	0.925	&	0.920	&	0.916		&	0.978	\\ \hline
			Unbalanced	&	5	&	0.981	&	0.004	&	0.989	&	0.981	&	0.984		&	0.999	\\ \hline
		\end{tabular}
	\end{center}
\vspace{-0.15cm}
\end{table}

\noindent\textbf{Metrics: }
To evaluate the performance of the classifier on the different classes and available features (F), standard machine learning metrics were used such as Precision (PR), Recall (RC), F-measure (FM), True Positive rate (TPR), False Positive rate (FPR), and area under the receiver operating curve (AUC).

\subsection{\csync\ in ID-sharing HTTP}\label{sec:id-sharing-vs-cs}

In this experiment, we assume there is already in place an existing technique for analysis of the HTTP traffic of the user, similar to the method outlined in Figure~\ref{fig:detection}.
However, there are candidate \csync\ events that cannot be confirmed, as the IDs cannot be matched with SET cookie IDs, either because these actions are not available to the method, or because the IDs are encrypted.

In this case, a machine learning classifier can be trained to detect if an id-sharing HTTP request is a true \csync\ event, by matching its pattern to past verified \csync\ events.
For this experiment, we use two classes: the \csync\ events and the \textit{id-sharing but non-\csync} events, to train and test a decision tree classifier under different subsets of features.
The training and testing was performed using 10 cross-fold validation process.
The results are shown in Table~\ref{tbl:ml-classifier-idsharing}.
We observe that independently, each of the features considered have some predictive power, except from the \textit{NoOfParams} feature.
When the most important features (using information gain as metric) are combined, a weighted AUC of 0.919 is achieved.
When we select non-ID related features, a weighted AUC=0.887 is reached, and with all features the classifier can reach a weighted AUC=0.946.

\subsection{\csync\ in HTTP with ID looking strings}\label{sec:random-vs-cs}
In this setup, we assume there is a simple HTTP pre-filter, keeping connections with ID looking strings for further investigation.
This is a necessary step to reduce the run-time workload of the classifier, as connections relevant to the task are only $\sim$$20\%$ of the overall HTTP workload.
Then, the classifier has to decide which of the 3 classes match for each of the selected HTTP requests:
1) \csync, 2) \textit{id-sharing but non-\csync}, 3) other.
In this case, \textit{other} refers to HTTP entries containing an ID-looking string, but are \textit{not id-sharing}.

We perform two rounds of tests on one month's data:
1) train and test the algorithm using balanced data from the three classes, in a 10 cross-fold validation process.
2) train the algorithm on balanced data from the three classes (as in (1)), but test it on an unseen and unbalanced dataset which maintains the ratio of the three classes: \csync: 1.6\%,  \textit{id-sharing but non-\csync}: 0.73\% other: 97.67\%.

As seen from the classification results (Table~\ref{tbl:ml-classifier-random}), the company name and parameter used are among the most important features; number of parameters is the worst.
Non-ID related features allow the classifier to reach weighted AUC = 0.973, with a high weighted Precision and Recall across all classes.
When all features are used, a weighted AUC = 0.978 is reached, similarly to the high importance feature set that disregards the number of parameters.
Interestingly, when the classifier is trained on the balanced dataset, and tested on the unbalanced test set (last row of Table~\ref{tbl:ml-classifier-random}), the classifier can distinguish very well the three classes, with low error rates across all three classes, even though there is high imbalance in the classes.
These results are further validated by the breakdown of performance per class, demonstrated in Table~\ref{tbl:ml-classifier-random-breakdown-unseen}, which show high TP rate and low FP rate for all three classes independently.

Overall, the results show that it is possible to understand and model the patterns of \csync, as they are driven by particular types of companies, using specific parameters, etc.
Therefore, an online classifier could be trained to provide insights as to what each HTTP connection is and how likely it is to be performing \csync, without the need to match the IDs to SET cookie actions.

\begin{table}[t]
	\begin{center}
		\footnotesize
		\caption{Detailed performance of decision tree model trained on different subsets of features in a balanced dataset, and tested on an unseen, and unbalanced test set, which maintains the original ratio of classes (last row of Table~\ref{tbl:ml-classifier-random}).
			C: CSync, ICS: \textit{id-sharing but non-\csync}, O: Other, WA: weighted average.
		}\vspace{-0.4cm}
		\label{tbl:ml-classifier-random-breakdown-unseen}
		\begin{tabular}{c|c|c|c|c|c|c }  \hline
			Class	&	TPR		&	FPR		&	PR		&	RC		&	FM		&	AUC \\ \hline
			CS		&	0.988	&	0.014	&	0.534	&	0.988	&	0.693	&	0.998	\\
			ICS		&	0.603	&	0.005	&	0.458	&	0.603	&	0.521	&	0.990	\\
			O		&	0.984	&	0.004	&	1.000	&	0.984	&	0.992	&	0.999	\\ \hline
			WA		&	0.981	&	0.004	&	0.989	&	0.981	&	0.984	&	0.999	\\ \hline
		\end{tabular}
	\end{center}
\vspace{-0.3cm}
\end{table}%

\vspace{-0.25cm}
\section{Related Work}
\label{sec:related}

\csync\ has become a commonplace on the Web.
One of the first works to discuss this mechanism~\cite{lukasz2014selling-privacy-auction} studies programmatic auctions from a privacy perspective and presents \csync\ as an integral part of communication between the participating entities.
The study identified over 100 \csync\ events while crawling the top 100 sites.
In our study, we extend their detection mechanism to detect \csync\ when cookieID is piggybacked in either URL parameters or path.

In~\cite{Acar:2014:WNF:2660267.2660347}, authors conduct a \csync\ privacy analysis by studying a small dataset of 3000 crawled sites, in conjunction with re-spawning cookies and how, together, they affect the reconstruction of user's browsing history by trackers.
In~\cite{www18adcost} they measure the advertising ecosystem cost to users.
Focusing on user privacy and targeted advertising, they use \csync\ as a metric for anonymity loss, showing that users receive 3.4 \csyncs\ per ad-impression.

Papadopoulos et al.~\cite{Papadopoulos:2018:ECM:3193111.3193117} present how \csync\ can wreck a secure browsing session.
They show cases where 3rd-parties may leak a user's cookie IDs and browsing history, thus increasing the identifiability of the user  to a snooping ISP. By probing the top 12k Alexa sites they find 1 out of 13 websites exposing their users to these privacy leaks even when they use TLS and secure VPN services. In a recent census by Englehardt et al.~\cite{Englehardt:2016:OTM:2976749.2978313}, authors measure \csync\ and its adoption in a small subset of 100,000 crawled sites, before highlighting the need of further investigation given its increased privacy implications.
Their results show that 157 of top 200 (i.e. 78\%) 3rd parties synchronize cookies with at least one other party.

In~\cite{Ghosh:2015:MME:2764902.2745801} they study the economics and the revenue implications of \csync\ from the point of view of an informed seller of advertising space, uncovering a trade-off between targeting and information leakage.
Similarly, in~\cite{bergemann2015selling},  authors explore the role of data providers on the price and allocation of consumer-level information and develop a simple model of data pricing that captures the key trade-offs involved in selling information encoded in 3rd-party cookies. 
In~\cite{Falahrastegar2016} they investigate tracking groups that share user-specific identifiers in a dataset collected after recording the browsing history of 100 users for two weeks.
In this dataset, they detect 660 ID-sharing groups and found domains with sensitive content (such as health-related) that shared IDs with domains related to ad-trackers.

In~\cite{bashir2016tracing} they aim to enhance the transparency in ad ecosystem with regards to information sharing, developing a content-agnostic methodology to detect client- and server-side flows of information between ad exchanges by leveraging retargeted ads. 
By using crawled data, authors collected 35448 ad impressions and identified 4 different kinds of information sharing behaviour between ad exchanges.
In~\cite{bashirPets}, they study the diffusion of user tracking caused by RTB-based programmatic ad-auctions, considering \csync\ as the core component of such auctions and the primary factor of the diffusion of privacy leaks. Results of their study show that under specific assumptions, no less than 52 tracking companies can observe at least 91\% of an average user's browsing history.

Contrary to these works, we conduct a first of its kind, full-scale study of \csync\ by analysing a year-long dataset of \totalUsers\ real users, thus avoiding any biases from crawling websites using artificial personas. Additionally, since nowadays mobile drives the majority of the network traffic, our work is the first to study \csync\ in the growing ecosystem of mobile devices.

\vspace{-0.15cm}
\section{Summary and Discussion}
\label{sec:discussion}
One of the most popular techniques for trackers to share the IDs they assign to users today is \cs, with which different domains can merge their databases in the background.
However, syncing userIDs of a given user increases the user identifiability while browsing, thus reducing their overall anonymity on the Web.
In this paper, we build \toolname: a holistic system to detect \csync\ events, either when the synced IDs are available in plaintext, or even when they are obfuscated (i.e., hashed, encrypted).
Using our detection mechanism, we are the first to explore \csync\ in the mobile ecosystem and the first to analyze it in depth, using a year-long dataset of real mobile users.
\toolname\ is able to capture 3.771\% more \csync\ cases than related work.

\noindent{\bf Results:}
Our analysis of $263k$ \csync\ requests, syncing $22.3k$ unique userIDs, led to the following findings:
\begin{itemize} [leftmargin=1.5em,topsep=0pt,itemsep=0pt]
\item 97\% of users are exposed to \csync\ at least once. The median user is synced at least once within the first week of browsing.
\item Ad-related domains participate in more than 75\% of all \csync\ through the year, learning as much as 90\% of all synced userIDs.
\item Three companies learn more than 30\% of all userIDs, each.
\item The median userID gets leaked to 3.5 domains, on average.
\item The average user receives around 1 synchronization per 68 HTTP requests, and gets up to 6.5 of their userIDs synced.
\item The number of domains that learn about the median user after \csyncs\ grows by a factor of 6.75.
\item We find 63 cases, where domains set cookies on the users' browsers with userIDs previously set by other domains. This universal identification model enables collaborating domains to share data without background database merges. 
\item We find 131 cases of domains storing in cookies their \csyncs\ results forming ID Summaries.
\item 5\% of users suffer from ID-spilling in their secure TLS traffic. 
\item several sensitive information (e.g., gender, birth dates) is passed to the syncing domain along with the userID.
\end{itemize}

In addition to the classic, heuristics-based method applied to collect the above findings, in this work, we proposed a novel, \csync\ detection mechanism able to detect at real time \csync\ events, even if the synced IDs are obfuscated.
In particular, this online, machine learning classifier can be trained to provide insights as to what each HTTP connection is, and how likely it is to be performing \csync, without the need to match the IDs to SET cookie actions.
We use the set of detected \csyncs\ from the heuristics-based method as ground truth, to train our machine learning, cookie-less detection algorithm.
We were able to achieve high accuracy (84\%-90\%) and high AUC (0.89-0.97), when non-ID related features were used.

\noindent{\bf Countermeasures:} Nowadays, the most popular defence mechanism of \csync\ is the use of the traditional ad-blockers. Indeed, since the vast majority (75\%) of \csync\ takes place among ad-related domains (see Figure~\ref{fig:IDsPerCat}), it is easy to anticipate that by blocking ad-related requests, one can eliminate a large portion of the privacy leak that \csync\ causes.
However, the all-out approach of ad-blockers causes significant harm on publishers' content monetization models, forcing some of them to deploy anti-adblocking mechanisms~\cite{mughees2017detecting,nithyanand2016adblocking,iqbal2017ad} and deny serving ad-blocking users~\cite{wired,forbes,financialTimes}. 

Mitigating mechanisms against \csync\ require a more targeted blocking strategy, that would not blindly harm the current ad-ecosystem.
Instead, by applying detection techniques such as \toolname, and blocking the specific traffic which has been found to facilitate \csync, we believe that the harmful privacy leakage and loss of anonymity of users due to \csync\ can be avoided, without the dire consequences on publishers' business models.

\noindent{\bf Impact:}
User data collection and sharing activities done without users' explicit consent can be illegal with hefty penalties imposed to companies involved, as described in the new EU regulations for protecting user personal data and online privacy (GDPR and e-Privacy).
Thus, it is important to design practical web transparency tools such as \toolname, readily available to privacy researchers, regulators and end-users to investigate personal data leakage and anonymity on the Web, due to 3rd parties' activities such as \csync~\cite{Iordanou:2018:TCB:3278532.3278561}.

\noindent{\bf ACKNOWLEDGEMENTS:} The research leading to these results has received funding from EU's Marie Sklodowska-Curie  and Horizon 2020 Research \& Innovation Programme under grant agreements 690972, 786669 and 830927. The paper reflects only the authors' view and the Commission is not responsible for any use that may be made of the information it contains.

\bibliographystyle{ACM-Reference-Format}
\balance
\bibliography{main}
\end{document}